\documentclass[10pt,aps,prd,reprint,floatfix,amsfonts,amssymb,amsmath,preprintnumbers,notitlepage,nobibnotes,nofootinbib]{revtex4-1}
\usepackage[ascii]{inputenc}
\usepackage[english]{babel}
\usepackage[Symbolsmallscale]{upgreek}
\usepackage{isomath}
\usepackage[protrusion=true,expansion=true,tracking=true,kerning=true,spacing=true,final,babel=true]{microtype}
\usepackage{physics}
\usepackage{empheq}
\usepackage{xcolor}
\usepackage{url}
\usepackage[final]{hyperref}

\begin{document}

\title{Anisotropies in the stochastic gravitational-wave background:\\Formalism and the cosmic string case}
\date{\today}
\author{Alexander~C.~Jenkins}
\email{Alexander.Jenkins@kcl.ac.uk}
\author{Mairi~Sakellariadou}
\email{Mairi.Sakellariadou@kcl.ac.uk}
\affiliation{Theoretical Particle Physics and Cosmology Group, Physics Department, King's College London, University of London, Strand, London WC2R 2LS, United Kingdom}
\preprint{KCL-PH-TH/2018-6}
\begin{abstract}
    We develop a powerful analytical formalism for calculating the energy density of the stochastic gravitational wave background, including a full description of its anisotropies.
    This is completely general, and can be applied to any astrophysical or cosmological source.
    As an example, we apply these tools to the case of a network of Nambu-Goto cosmic strings.
    We find that the angular spectrum of the anisotropies is relatively insensitive to the choice of model for the string network, but very sensitive to the value of the string tension $G\mu$.
\end{abstract}
\maketitle

\section{Introduction}
The direct detection of gravitational waves (GW) from binary black hole mergers~\cite{Abbott:2016blz,Abbott:2016nmj,Abbott:2017vtc,Abbott:2017gyy,Abbott:2017oio} and from a binary neutron star merger~\cite{TheLIGOScientific:2017qsa} has opened a new window to the Universe.
Gravitational waves offer a powerful tool for understanding the early stages of the Universe, particularly the prerecombination era that is inaccessible to conventional (electromagnetic) astronomy.
Apart from the events so far detected by the LIGO and Virgo collaborations, we expect many more which are too distant to be individually detected.
These quieter events, produced by many weak, independent and unresolved sources, constitute the stochastic GW background (SGWB).
A variety of sources may lead to a SGWB, such as compact binary mergers, cosmic strings~\cite{Vachaspati:1984gt,Sakellariadou:1990ne} or phase transitions in the early Universe~\cite{Binetruy:2012ze}, while at much higher redshifts one expects a contribution from a cosmological background due to a mechanism such as inflation.

Gravitational wave sources with an inhomogeneous spatial distribution lead to a SGWB characterized by preferred directions, and hence anisotropies.
The main contribution to such an anisotropic background comes from astrophysical sources (such as compact binaries) that follow the local distribution of matter.
The finiteness of the GW sources and the nature of the spacetime along the line of propagation of GWs will also contribute to anisotropies in the SGWB.
The aim of this work is to develop a formalism for anisotropies in the SGWB of any astrophysical or cosmological source, and then apply it to the case of GWs sourced by cosmic string networks.

Our study is divided into two parts.
In Sec.~\ref{sec:general-formalism}, we follow the formalism presented in Ref.~\cite{Cusin:2017fwz}, which we develop further in order to derive a general expression for anisotropies in the SGWB, written in a form consistent with the usual GW literature.
In addition, we derive a simple condition for the SGWB to be a Gaussian random field (GRF), and make a clear distinction between background and foreground sources in order to calculate the background in an unbiased way.
We compute the kinematic dipole, which must be subtracted since it interferes with the anisotropy statistics.
Finally, we show how to relate our results to future observational work.
In Sec.~\ref{sec:cosmic-strings}, we apply this formalism to the case of cosmic string networks.
In particular, we study gravitational waves emitted from cusps, kinks and kink-kink collisions for three analytic models of Nambu-Goto string networks~\cite{Vilenkin:2000jqa,Blanco-Pillado:2013qja,Lorenz:2010sm}.

\section{General formalism}
\label{sec:general-formalism}
Consider a Friedman-Lema\^{\i}tre-Robertson-Walker (FLRW) spacetime with scalar perturbations,
    \begin{equation}
        \dd{s^2}=a^2\qty[-\qty(1+2\psi)\dd{\eta^2}+\qty(1-2\phi)\dd{\vb*x}\vdot\dd{\vb*x}],
    \end{equation}
    where $a\qty(\eta)$ is the scale factor, $\eta$ denotes conformal time and $\psi\qty(\eta,\vb*x)$, $\phi\qty(\eta,\vb*x)$ are the two Bardeen potentials, decomposed as $\psi=\Psi+\Pi$, $\phi=\Psi-\Pi$ respectively.
Using units with $c=\hbar=1$, setting $a\qty(\eta_\mathrm{o})=1$, and keeping only linear order perturbations, the energy density of GWs with observed frequency $\nu_\mathrm{o}$ arriving from a solid angle $\sigma_\mathrm{o}$ centered on the direction $\vu*e_\mathrm{o}$, is given in Ref.~\cite{Cusin:2017fwz} as
    \begin{align}
    \begin{split}
        \label{eq:cusin-et-al-main-result}
        &\frac{\dd[3]{\rho_\mathrm{gw}}}{\dd{\nu_\mathrm{o}\dd[2]{\sigma_\mathrm{o}}}}\qty(\nu_\mathrm{o},\vu*e_\mathrm{o})=\\
        &\qquad\frac{1}{4\uppi}\int_0^{\eta_\mathrm{o}}\dd{\eta}a^4\int\dd{\vb*\zeta}\bar{n}\mathcal{L}_\mathrm{s}\bigg[1+\delta_n-3\qty(\Psi_\mathrm{o}+\Pi_\mathrm{o})\\
        &\qquad\quad+4\qty(\Psi+\Pi)+\vu*e_\mathrm{o}\vdot\qty(3\vb*v_\mathrm{o}-2\vb*v)+6\int_\eta^{\eta_\mathrm{o}}\dd{\eta'}\pdv{\Psi}{\eta'}\bigg],
    \end{split}
    \end{align}
    with ``s" and ``o" subscripts indicating quantities evaluated at the GW source and at the observer, respectively, and with the $\eta$ integral along the line of sight, $\vb*x\qty(\eta,\vu*e_\mathrm{o})=\vb*x_\mathrm{o}+\qty(\eta_\mathrm{o}-\eta)\vu*e_\mathrm{o}$.
Note that $\vb*v\qty(\eta,\vu*e_\mathrm{o})$ stands for the peculiar 3-velocity of the cosmic fluid.
Here, $\mathcal{L}_\mathrm{s}\qty(\nu_\mathrm{s},\vb*\zeta)$ is the gravitational luminosity at emitted frequency $\nu_\mathrm{s}$ of a source with parameters $\vb*\zeta$, with the emitted frequency given in terms of the observed frequency $\nu_\mathrm{o}$ by
    \begin{align}
    \begin{split}
        \label{eq:source-frequency}
        \nu_\mathrm{s}=\frac{\nu_\mathrm{o}}{a}\bigg[1&+\Psi_\mathrm{o}+\Pi_\mathrm{o}-\Psi-\Pi\\
        &+\vu*e_\mathrm{o}\vdot\qty(\vb*v-\vb*v_\mathrm{o})-2\int_\eta^{\eta_\mathrm{o}}\dd{\eta'}\pdv{\Psi}{\eta'}\bigg].
    \end{split}
    \end{align}
We also define $n\qty(\eta,\vu*e_\mathrm{o},\vb*\zeta)$ as the source number density---per \emph{physical} volume, not comoving volume---with homogeneous background value $\bar{n}\qty(\eta,\vb*\zeta)$.
The number density inhomogeneities are expressed in terms of the density contrast
    \begin{equation}
        \delta_n\qty(\eta,\vu*e_\mathrm{o},\vb*\zeta)\equiv\frac{n-\bar{n}}{\bar{n}},
    \end{equation}
    so that $n=\bar{n}\qty(1+\delta_n)$.

In order to express Eq.~\eqref{eq:cusin-et-al-main-result} in a form consistent with the SGWB literature, we change from linear to logarithmic frequency, and normalize with respect to the critical density $\rho_{\mathrm{c}}=3H_0^2/(8\uppi G)$, giving the \emph{density parameter},
    \begin{equation}
        \Omega_\mathrm{gw}\qty(\nu_\mathrm{o},\vu*e_\mathrm{o})\equiv\frac{1}{\rho_\mathrm{c}}\frac{\dd[3]{\rho_\mathrm{gw}}}{\dd(\ln\nu_\mathrm{o})\dd[2]{\sigma_\mathrm{o}}}
        =\frac{8\uppi G\nu_\mathrm{o}}{3H_\mathrm{o}^2}\frac{\dd[3]{\rho_\mathrm{gw}}}{\dd{\nu_\mathrm{o}}\dd[2]{\sigma_\mathrm{o}}}.
    \end{equation}
Thus, using the above definition, the dimensionless quantity expressing the intensity of a stochastic background of gravitational waves, in the context of a FLRW universe with scalar perturbations, is
    \begin{align}
    \begin{split}
    \label{eq:Omega-gw}
        &\Omega_\mathrm{gw}\qty(\nu_\mathrm{o},\vu*e_\mathrm{o})\\
        &=\frac{2G\nu_\mathrm{o}}{3H_\mathrm{o}^2}\int_0^{\eta_\mathrm{o}}\dd{\eta}a^4\int\dd{\vb*\zeta}\bar{n}\mathcal{L}_\mathrm{s}\bigg[1+\delta_n-3\qty(\Psi_\mathrm{o}+\Pi_\mathrm{o})\\
        &\qquad+4\qty(\Psi+\Pi)+\vu*e_\mathrm{o}\vdot\qty(3\vb*v_\mathrm{o}-2\vb*v)+6\int_\eta^{\eta_\mathrm{o}}\dd{\eta'}\pdv{\Psi}{\eta'}\bigg].
    \end{split}
    \end{align}
We decompose this in terms of the isotropic monopole term $\bar{\Omega}_\mathrm{gw}\qty(\nu_\mathrm{o})$ and the GW energy density contrast $\delta_\mathrm{gw}\qty(\nu_\mathrm{o},\vu*e_\mathrm{o})$, giving
    \begin{equation}
        \label{eq:monopole-density-contrast}
        \Omega_\mathrm{gw}\equiv\bar{\Omega}_\mathrm{gw}\qty(1+\delta_\mathrm{gw}).
    \end{equation}
This definition implicitly takes the average of $\delta_\mathrm{gw}$ over the celestial sphere as zero, so we must choose a gauge in which the spatial average of the cosmological potentials is also zero (the spatial average of the density contrast $\delta_n$ is zero by definition).
Note that $\bar{\Omega}_\mathrm{gw}$ corresponds to the average GW flux at frequency $\nu_\mathrm{o}$ \emph{per unit solid angle}, so that the total flux at this frequency is $4\uppi\bar{\Omega}_\mathrm{gw}$.
This factor of $4\uppi$ must be taken into account when comparing our results with isotropic models of the SGWB, as the latter are usually expressed in terms of the total flux.

\subsection{Relating strain and luminosity}
\label{sec:relating-strain-luminosity}
The gravitational luminosity $\mathcal{L}_\mathrm{s}$ of any astrophysical or cosmological source that emits a series of GW signals can be decomposed as
    \begin{equation}
        \label{eq:luminosity-spectrum}
        \mathcal{L}_\mathrm{s}\qty(\nu_\mathrm{s},\vb*\zeta)=\dv{E_\mathrm{s}}{\nu_\mathrm{s}}R\qty(\vb*\zeta),
    \end{equation}
    where $E_\mathrm{s}\qty(\vb*\zeta)$ is the total energy lost from the system due to each signal, and $R\qty(\vb*\zeta)$ is the rate at which the signals are emitted (i.e.~the product $nR$ is the signal rate per unit physical volume, per unit source time).\footnote{Note that Eq.~\eqref{eq:luminosity-spectrum} is valid regardless of the signal duration. For cosmic strings, the duration of the signal is typically much shorter than the period between signals $1/R$ (since the duration goes as the inverse of the frequency $\nu$, the rate goes as the inverse of the loop length $l$, and we are interested in higher harmonics $\nu\gg l$). However, one could also apply Eq.~\eqref{eq:luminosity-spectrum} to extremely long-duration signals, such as the ``continuous waves" produced by the quadrupole moment of a rotating neutron star. In this case, $\dv*{E}{\nu}$ would be interpreted as the energy spectrum for some time interval $T\gg1/\nu$, and $R$ would simply be $1/T$. Since the total energy emitted in time $T$ is proportional to $T$ for continuous sources, this gives an unambiguous value for the luminosity spectrum $\mathcal{L}_\mathrm{s}$.}
We compute $E_\mathrm{s}$ as a function of the GW strain $h_{\mu\nu}$ by integrating the solid angle $\dd[2]{\sigma_\mathrm{s}}$ over a spherical surface of radius $r_\mathrm{s}$ centered on the source, where $r_\mathrm{s}$ is large enough to use linearized general relativity on a Minkowski background, but small enough to neglect cosmological effects.
We thus obtain
    \begin{align*}
        E_\mathrm{s}&=\frac{1}{32\uppi G}\int_{S^2}\dd[2]{\sigma_\mathrm{s}}r\mathrlap{^2}_\mathrm{s}\int_{-\infty}^{+\infty}\dd{t_\mathrm{s}}\pdv{h_{ij}^\mathrm{TT}}{t_\mathrm{s}}\pdv{h_{ij}^\mathrm{TT}}{t_\mathrm{s}}\\
        &=\frac{1}{32\uppi G}\int_{S^2}\dd[2]{\sigma_\mathrm{s}}r\mathrlap{^2}_\mathrm{s}\int_{-\infty}^{+\infty}\dd{t_\mathrm{s}}\sum_{A=+,\times}\left(\pdv{h_A}{t_\mathrm{s}}\right)^2,
    \end{align*}
    where $h_{ij}^\mathrm{TT}\qty(t_\mathrm{s},\vb*x_\mathrm{s},\vb*\zeta)$ is the strain in the transverse traceless (TT) gauge, with ``plus" and ``cross" mode amplitudes $h_+,h_\times$, and $\qty(t_\mathrm{s},\vb*x_\mathrm{s})$ are the co\"ordinates of the local Minkowski metric~\cite{Maggiore:1900zz}.
Writing the strain $ h_A$ in terms of its Fourier transform $\tilde{h}_A$,
    \begin{equation*}
        h_A\qty(t_\mathrm{s})=\int_{-\infty}^{+\infty}\dd{\nu_\mathrm{s}}\mathrm{e}^{2\uppi\mathrm{i}\nu_\mathrm{s}t_\mathrm{s}}\tilde{h}_A\qty(\nu_\mathrm{s}),
    \end{equation*}
    and using $\tilde{h}_A\qty(-\nu_\mathrm{s})=\tilde{h}^*_A\qty(\nu_\mathrm{s})$ (since $h_A$ is real), we find
    \begin{equation*}
        E_\mathrm{s}=\frac{\uppi }{4G}\int_{S^2}\dd[2]{\sigma_\mathrm{s}}r\mathrlap{^2}_\mathrm{s}\int_0^{+\infty}\dd{\nu_\mathrm{s}}\nu_\mathrm{s}^2\sum_{A=+,\times}\qty|\tilde{h}_A\qty(\nu_\mathrm{s})|^2,
    \end{equation*}
    which we have written in terms of positive frequencies only (i.e. this is a one-sided spectrum).
Since in what follows we are not interested in polarization effects, we can simplify the above expression by defining the total strain magnitude
    \begin{equation}
        \tilde{h}\equiv\sqrt{\frac{|\tilde{h}_+|^2+|\tilde{h}_\times|^2}{2}}.
    \end{equation}
Rewriting $E_\mathrm{s}$ in terms of $ \tilde{h}$ and using the definition in Eq.~\eqref{eq:luminosity-spectrum}, the luminosity spectrum of a single source is therefore given by
    \begin{equation}
        \label{eq:luminosity-strain}
        \mathcal{L}_\mathrm{s}\qty(\nu_\mathrm{s},\vb*\zeta)=\frac{\uppi\nu^2_\mathrm{s}R\qty(\vb*\zeta)}{2G}\int_{S^2}\dd[2]{\sigma_\mathrm{s}}r\mathrlap{^2}_\mathrm{s}\tilde{h}^2.
    \end{equation}
Using Eq.~\eqref{eq:Omega-gw} and Eq.~\eqref{eq:luminosity-strain}, the density parameter $\Omega_\mathrm{gw}$ is given, to linear order, by
    \begin{align}
    \begin{split}
        \label{eq:omega-gw}
        &\Omega_\mathrm{gw}\qty(\nu_\mathrm{o},\vu*e_\mathrm{o})\\
        &=\frac{\uppi\nu_\mathrm{o}^3}{3H_\mathrm{o}^2}\int_0^{\eta_\mathrm{o}}\dd{\eta}a^2\int\dd{\vb*\zeta}\bar{n}R\bigg[1+\delta_n-\Psi_\mathrm{o}-\Pi_\mathrm{o}\\
        &\qquad\qquad+2\qty(\Psi+\Pi)+\vu*e_\mathrm{o}\vdot\vb*v_\mathrm{o}+2\int_\eta^{\eta_\mathrm{o}}\dd{\eta'}\pdv{\Psi}{\eta'}\bigg]\\
        &\qquad\qquad\qquad\qquad\times\int_{S^2}\dd[2]{\sigma_\mathrm{s}}r\mathrlap{^2}_\mathrm{s}\tilde{h}^2,
    \end{split}
    \end{align}
    where we have used Eq.~\eqref{eq:source-frequency} to convert the source-frame frequency in Eq.~\eqref{eq:luminosity-strain} to the corresponding observer-frame frequency.
Note that this changes the linear perturbation terms---in particular, there is no net contribution from the source's peculiar motion, only from that of the observer.

\subsection{Gaussian and non-Gaussian backgrounds}
\label{sec:gaussian-non-gaussian}
Analyzing the anisotropic statistics of the SGWB is greatly simplified if $\Omega_\mathrm{gw}$ is a Gaussian random field (GRF).
In particular, Wick's theorem tells us that if the field is Gaussian, we can \emph{fully} characterize its anisotropies in terms of the mean $\bar{\Omega}_\mathrm{gw}$ and the two-point correlation function (2PCF), as defined in Sec.~\ref{sec:formalism-sgwb-decomposition}.
It is also convenient from the point of view of GW data analysis if the detector strain $h\qty(t)$ associated with the SGWB is a Gaussian process, as this gives a simple likelihood function for the strain~\cite{Romano:2016dpx}.
Current LIGO/Virgo searches for the SGWB exploit this fact, and use pipelines optimized for Gaussian backgrounds, though we note that search methods for non-Gaussian backgrounds do exist (see e.g.~Ref.~\cite{Smith:2017vfk}).

However, one must be careful when speaking of a ``Gaussian background", as it is not clear \emph{a priori} that $h\qty(t)$ being a Gaussian process implies that $\Omega_\mathrm{gw}$ is a GRF, or vice versa.
In this section we use a simple model of a background composed of independent GW bursts to derive sufficiency conditions for Gaussianity of each of the relevant quantities: first, we reproduce the standard condition that gives a Gaussian strain $h\qty(t)$; then we find a \emph{different} condition that makes the isotropic energy density $\bar{\Omega}_\mathrm{gw}\qty(\nu_\text{o})$ Gaussian; and finally we extend this to find a condition for the energy density as a function of sky location $\Omega_\mathrm{gw}\qty(\nu_\text{o},\vu*e_\text{o})$ to be a GRF, given some angular resolution $\updelta\sigma$.

\subsubsection{A simple model of an incoherent SGWB}
Suppose we observe the SGWB over a time interval $T$.
It can then be written in terms of a discrete set of frequencies $\nu=n/T$, where $n\in\mathbb{Z}_{>0}$.
Let us focus on the signal in a single frequency bin centered on $\nu$ (with width $\upDelta\nu=1/T$ set by the observation time).
Since we are considering a background composed of many independent transient bursts, we write the complex GW strain at this frequency $h\qty(t)\equiv h_+\qty(t)-\mathrm{i}h_\times\qty(t)$ as the sum of $N$ bursts
    \begin{align}
    \begin{split}
        \label{eq:sgwb-bursts}
        h\qty(t)&=\sum_{i=1}^Nh_i\qty(t),\\
        h_i\qty(t)&\equiv
        \begin{cases}
            A_i\mathrm{e}^{\mathrm{i}\qty(2\uppi \nu t+\phi_i)} & \text{if }t_i\le t\le t_i+\upDelta t,\\
            0 & \text{else},
        \end{cases}
    \end{split}
    \end{align}
    where the (real) amplitudes $A_i$, times of arrival $t_i$, and initial phases $\phi_i$ of the bursts are all random variables.
We take the amplitudes as being independent and identically distributed (i.i.d.) according to some unknown probability distribution that depends on the frequency bin.
The times of arrival are distributed according to a Poisson process with rate parameter $R$ (also dependent on frequency bin), while the phases are uniformly distributed on $[0,2\uppi)$.

When we speak of ``burst signals", we mean signals whose duration $\upDelta t$ is ``short" in some sense.
We can quantify this by saying that a burst lasts for no more than $\order{1}$ wavelengths in each frequency bin, so that its duration in a given frequency bin can be taken as $\upDelta t\approx1/\nu$.
This is a good approximation for most burstlike signals (e.g. supernovae~\cite{Buonanno:2004tp} or cosmic string cusps and kinks~\cite{Damour:2001bk}).
Specific GW sources will have different signal durations, but for the sources mentioned above, and for more general transient sources, this approximation is accurate to within an order of magnitude.
This is \emph{not} the case for GW signals from compact binary coalescences, where the duration in some small frequency interval $\qty[\nu,\nu+\updelta\nu]$ is roughly $\upDelta t\approx5\updelta\nu/\qty(96\uppi^{8/3}\mathcal{M}^{5/3}\nu^{11/3})$ during the inspiral phase (where $\mathcal{M}$ is the chirp mass)~\cite{Maggiore:1900zz}.
For a discussion of the Gaussianity of the stochastic background in this case, see Ref.~\cite{Jenkins:2018uac}.

\subsubsection{Conditions for \texorpdfstring{$h\qty(t)$}{} to be Gaussian}
It is well known that for $h\qty(t)$ to be a Gaussian process, it is sufficient for the \emph{duty cycle} to be much greater than unity.
We define this quantity below and give a brief justification of this condition, using the simple model described above.

At any time $t$, the observed GW strain due to the SGWB is the superposition of all the bursts $h_i$ with arrival times up to $1/\nu$ before the time $t$, as each burst has a duration of $1/\nu$.
This means that $h\qty(t)$ is the sum of some number of i.i.d.~random variables, and in the limit where this number is large $h\qty(t)$ is Gaussian by the central limit theorem.

Since the times of arrival are given by a Poisson process with rate $R$, the total number of bursts $N$ will tend to $RT$ in the limit where $T\gg1/R$.
(The \emph{expected} number of bursts will always be $RT$. However, there will be random fluctuations around this value, which vanish only when $RT\to\infty$.)
For each of these $N$ bursts, there is a probability of roughly $1/\nu T$ that they will arrive at a time between $t-1/\nu$ and $t$, so the \emph{expected} number of bursts contributing to the strain at time $t$ is $N/\nu T$.
By the law of large numbers, the number of contributing bursts therefore converges to $N/\nu T$ in the limit where $N\gg1$.
So for $RT\gg1$, the number of bursts in-band at time $t$ converges to $R/\nu$.
This quantity is called the \emph{duty cycle}, $\Lambda\equiv R\upDelta t\approx R/\nu$.
In order to ensure that $h\qty(t)$ is Gaussian, it is therefore sufficient to take $RT\gg1$ and $\Lambda\gg1$.
In these limits, the fluctuations in the number of signals with respect to time are small, so if the signal is Gaussian at some time $t$ then it is Gaussian for the whole observing period $T$.

For reasons discussed below, we usually only consider frequencies $\nu\gg1/T$, so the limit $R\gg\nu$ implies that $RT\gg1$.
We are therefore left with a single sufficiency condition for Gaussianity:
    \begin{equation}
        \Lambda\gg1\quad\Longrightarrow\quad\forall t\in[0,T],\quad h\qty(t)\text{ is Gaussian}.
    \end{equation}
When ``Gaussian" GW backgrounds are discussed in the literature, this is usually what is meant.
For studying anisotropies in the background, however, it is the density parameter $\Omega_\mathrm{gw}$ that is important, rather than the strain.

\subsubsection{Conditions for \texorpdfstring{$\bar{\Omega}_\mathrm{gw}$}{} to be Gaussian}
As we hinted at before, $h\qty(t)$ being Gaussian at frequency $\nu$ is \emph{not} the same as the isotropic energy density $\bar{\Omega}_\mathrm{gw}\qty(\nu)$ being Gaussian.
To see this, we express $\bar{\Omega}_\mathrm{gw}$ explicitly using
    \begin{equation}
        \bar{\Omega}_\mathrm{gw}=\frac{\nu}{\upDelta\nu}\frac{1}{48\uppi H_\mathrm{o}^2}\ev{\dot{h}\dot{h}^*}.
    \end{equation}
Here $\upDelta\nu=1/T$ is the frequency resolution.
The factor of $\nu/\upDelta\nu$ is equivalent to the derivative $\dv{}{\qty(\ln\nu)}=\nu\dv{}{\nu}$ used in the continuum case $T\to\infty$.
The angle brackets represent an average over many periods, as this is required to have a well-defined notion of ``energy" for a GW.
It is only possible to perform this average if we observe the SGWB for many periods, so we must have $T\gg1/\nu$.
Assuming this is the case, we use the decomposition Eq.~\eqref{eq:sgwb-bursts} to find
    \begin{equation}
        \label{eq:sgwb-amplitude-explicit}
        \bar{\Omega}_\mathrm{gw}=\frac{\uppi\nu^2}{12H_\mathrm{o}^2}\qty[\sum_{i=1}^NA^2_i+\sum_{\substack{\text{coincident}\\\text{pairs }\{i,j\}}}B_{ij}],
    \end{equation}
    where $B_{ij}\equiv A_iA_j\qty(1-\nu\qty|t_i-t_j|)\cos\qty(\phi_i-\phi_j)$.
As well as the contribution due to the energy of each individual burst (the first sum in the expression above), we also have a contribution from cross-terms between coincident bursts (the second sum), whose times of arrival $t_i,t_j$ are within $1/\nu$ of each other.
There are $N^2-N$ pairs of bursts, and probability of any pair of bursts overlapping is roughly $1/\nu T$, so by the law of large numbers, the number of coinciding pairs converges to $N\qty(N-1)/\nu T$ when $N\qty(N-1)\gg1$.
As before, taking $RT\gg1$ ensures that $N\to RT$.
The random variables $A^2_i$ and $B_{ij}$ are i.i.d.~for all bursts $i$ and for all coincident pairs $\{i,j\}$ respectively, so the central limit theorem guarantees that $\bar{\Omega}_\mathrm{gw}$ is Gaussian if $N\gg1$ and $N\qty(N-1)/\nu T\gg1$.

Putting all this together, we find that the conditions
    \begin{equation}
        RT\gg1,\qquad RT\qty(RT-1)/\nu T\gg1,
    \end{equation}
    are sufficient for $\bar{\Omega}_\mathrm{gw}$ to be Gaussian at frequency $\nu$.
With some rearranging, we see that the second condition implies the first.
Rewriting in terms of the duty cycle, we have
    \begin{equation}
        \nu T\gg\frac{1}{\Lambda}+\frac{1}{\Lambda^2}\quad\Longrightarrow\quad\bar{\Omega}_\mathrm{gw}\qty(\nu)\text{ is Gaussian},
    \end{equation}
    where we only consider frequencies $\nu\gg1/T$.

We see that $\bar{\Omega}_\mathrm{gw}$ is always Gaussian if $\Lambda\ge1$.
This means that $h\qty(t)$ being Gaussian implies that $\bar{\Omega}_\mathrm{gw}$ is Gaussian, but note that the converse does not hold.
In fact, no matter how non-Gaussian $h\qty(t)$ is (i.e. no matter how small the duty cycle is), it is in principal possible to make $\bar{\Omega}_\mathrm{gw}$ Gaussian by increasing the observation time $T$ (the required observation time will depend entirely upon the duty cycle of the sources considered).

\subsubsection{Conditions for \texorpdfstring{$\Omega_\mathrm{gw}\qty(\nu,\vu*e)$}{} to be a GRF}
The discussion thus far has been about the \emph{isotropic} GW energy density, $\bar{\Omega}_\text{gw}$.
To extend this to the angular distribution of this energy as a field on the sky, we divide the sphere into some number of pixels $N_\mathrm{pix}$ of equal size $\updelta\sigma$, and let $\Omega_{\mathrm{gw},i}$ be the energy density in GWs arriving from the $i\mathrm{th}$ pixel.
If the background is statistically isotropic, then the probability of a given burst arriving from one particular pixel is $1/N_\mathrm{pix}=\updelta\sigma/4\uppi$.
If the number of bursts $N$ is large, then the number arriving from the $i$-th pixel converges to $N/N_\mathrm{pix}$.
Referring back to our discussion about Eq.~\eqref{eq:sgwb-amplitude-explicit}, we see that for $RT\gg1$ the number of bursts in a given pixel converges to $RT/N_\mathrm{pix}$, and the number of coincident pairs of bursts converges to $\frac{RT}{N_\mathrm{pix}}\qty(\frac{RT}{N_\mathrm{pix}}-1)/\nu T$.
For the total energy in that pixel from each burst and from cross-terms to be Gaussian, it is therefore sufficient to have
    \begin{equation}
        RT\gg1,\qquad\frac{1}{\nu T}\frac{RT}{N_\mathrm{pix}}\qty(\frac{RT}{N_\mathrm{pix}}-1)\gg1.
    \end{equation}
If this is the case, then \emph{all} the pixels are Gaussian, and the SGWB is a Gaussian random field.
As before, the second condition above implies the first, so we simplify to find that
    \begin{equation}
        \nu T\gg\frac{N_\mathrm{pix}}{\Lambda}+\frac{N_\mathrm{pix}^2}{\Lambda^2}\quad\Longrightarrow\quad\Omega_\mathrm{gw}\qty(\nu,\vu*e_\mathrm{o})\text{ is a GRF},
    \end{equation}
    where we emphasize once more that we are only interested in frequencies $\nu\gg1/T$.
We can also eliminate $N_\mathrm{pix}$ in favour of the angular resolution $\updelta\sigma$ to write this as
    \begin{equation}
        \label{eq:grf-condition}
        \nu T\gg\frac{4\uppi}{\Lambda\updelta\sigma}+\frac{16\uppi^2}{\Lambda^2\qty(\updelta\sigma)^2}\quad\Longrightarrow\quad\Omega_\mathrm{gw}\qty(\nu,\vu*e_\mathrm{o})\text{ is a GRF}.
    \end{equation}
In practice, we expect that it is only necessary for the LHS to be an order of magnitude larger than the RHS.

Equation~\eqref{eq:grf-condition} could potentially be a useful guide for the future observing strategies of advanced GW detectors.
Given an estimate of the duty cycle $\Lambda$ of a particular background source in a given frequency bin, and given the angular resolution $\updelta\sigma$ of the detector network, Eq.~\eqref{eq:grf-condition} specifies the requisite observing time $T$ to ensure that the field is Gaussian.
(Note that this time need not correspond to one unbroken observing period; it will likely be necessary to combine many separate observing runs.)
In principle, any background can be made to satisfy the criterion Eq.~\eqref{eq:grf-condition} at any angular resolution by increasing $T$, so our treatment in Sec.~\ref{sec:formalism-sgwb-decomposition} and~\ref{sec:observations} assumes that this criterion is met.
In practice, it may be desirable to measure the integral of $\Omega_\mathrm{gw}$ over some frequency interval much larger than the frequency bin size $1/T$, as this would increase the integrated GW power and therefore make Gaussianity more achievable for shorter observing times.

We emphasize once again that Eq.~\eqref{eq:grf-condition} is only relevant for a stochastic background composed of GW bursts that decay after $\order{1}$ wavelengths in-band, such that their duration can be approximated by $\upDelta t\approx1/\nu$.
In particular, it does \emph{not} apply to the astrophysical background from compact binaries, due to the assumption about the GW burst duration---this case is addressed in Ref.~\cite{Jenkins:2018uac}.
It also does not apply to a background from continuous sources (or very long transients, lasting longer than the observation time).
However, this case is somewhat simpler as there is a fixed number of continuous signals $N$, whose distribution amongst the pixels does not vary with time.
By a very similar argument to that given above, having $N\gg1$ continuous sources will ensure $h\qty(t)$ is Gaussian (by the central limit theorem), and having $N/N_\mathrm{pix}\gg1$ and $\tfrac{N}{N_\mathrm{pix}}\qty(\tfrac{N}{N_\mathrm{pix}}-1)\gg1$ ensures that $\Omega_\mathrm{gw}$ is a GRF (as the number of signals and number of overlapping signals per pixel are then large enough for the central limit theorem to apply).

\subsection{Separating background from foreground}
Equation~\eqref{eq:omega-gw} includes \emph{all} of the GW sources considered as part of the stochastic background.
However, to calculate the SGWB in an unbiased way, one must be careful not to include any loud, rare, individually resolvable signals that make up the \emph{foreground}~\footnote{The word ``foreground" is often used in the GW literature to describe ``nuisance" signals that obscure the source(s) of interest. Here, we use ``foreground" to mean any GW sources that are not part of the stochastic background.}---this was pointed out for the case of cosmic strings in Ref.~\cite{Damour:2001bk}.

There has been some debate in the literature over what constitutes a ``resolvable" signal.
Arguably the most thorough approach is to decide this on a signal-by-signal basis with Bayesian model selection, as in Ref.~\cite{Cornish:2015pda}.
For our purposes however, it is sufficient to distinguish between the two using the duty cycle $\Lambda\qty(\nu_\mathrm{o})$.
As above, this is defined as the average number of overlapping signals at frequency $\nu_\mathrm{o}$ experienced by the observer~\cite{Regimbau:2011bm}.
For foreground signals we have $\Lambda\ll1$, as the majority of the observation time contains no such signals (equivalently, the typical interval between these signals arriving is much greater than their duration).
For the SGWB we have $\Lambda\gg1$, as this background consists of a large number of superimposed signals (equivalently, the interval between signals that are part of the background is much shorter than their duration).
We stress that this is a \emph{detector-independent} (and therefore more general) way of defining what we mean by the ``stochastic background".
There will be many GWs that are not resolved by the detector network but which have $\Lambda\ll1$, and therefore could in principle be resolved by an idealized zero-noise detector; these might reasonably be described as ``background signals", but here we consider them part of the foreground.

Let $\Lambda\qty(\nu_\mathrm{o},\eta)$ denote the duty cycle for observed signals that are emitted from conformal time $\eta$ onward---i.e.~the conformal time at emission $\eta_\mathrm{s}$ obeys $\eta\le\eta_\mathrm{s}\le\eta_\mathrm{o}$.
Then we define the SGWB as all of the signals emitted at times $\eta_\mathrm{s}\le\eta_*$, where $\eta_*\qty(\nu_\mathrm{o})\le\eta_\mathrm{o}$ is defined by
    \begin{equation}
        \Lambda\qty(\nu_\mathrm{o},\eta_*)\equiv1.
    \end{equation}
We are thus excluding nearby sources whose combined duty cycle is less than unity, meaning that, on average, they do not overlap in time.
The SGWB is what remains: a continuous signal composed of many objects at large distances $\eta>\eta_*$.
If for a given frequency $\nu_\mathrm{o}$ there is no solution to the above equation, then we let $\eta_*(\nu_\mathrm{o})=0$; this means that there are not enough sources at this frequency to constitute a background.
Since we compute the duty cycle as an average quantity, we take $\eta_*$ as being the same in all directions on the sky.

Note that the duty cycle used in Sec.~\ref{sec:gaussian-non-gaussian} includes all signals that are part of the background, which is equal to the total duty cycle of all (background \emph{and} foreground) signals, $\Lambda_\text{tot}\equiv\Lambda\qty(\eta=0)$, minus the duty cycle of foreground signals, $\Lambda\qty(\eta_*)=1$.
When checking the Gaussianity of the background, the appropriate $\Lambda$ to use in Eq.~\eqref{eq:grf-condition} is therefore $\Lambda_\text{tot}-1$.

In order to calculate $\eta_*$, we write
    \begin{equation}
        \Lambda\qty(\nu_\mathrm{o},\eta_*)=\upDelta t\int\dd{\vb*\zeta}\int_{\eta_*}^{\eta_\mathrm{o}}\dd[3]{V\qty(\eta)}f_\mathrm{o}\bar{n}R,
    \end{equation}
    where we define $f_\mathrm{o}\qty(\nu_\mathrm{o},\eta,\vu*e_\mathrm{o},\vb*\zeta)$ as the fraction of the emitted signals that are observable at frequency $\nu_\mathrm{o}$ (this accounts for e.g.~beaming effects and cutoffs in the frequency spectrum of the signal).
So, $\Lambda\qty(\nu_\mathrm{o},\eta_*)$ is just the rate of arrival of observable signals originating at $\eta_\mathrm{s}\ge\eta_*$, multiplied by their duration $\upDelta t$.
In principle we should allow $\upDelta t$ to depend on $\eta$ and $\vb*\zeta$, but for burst signals such as those we consider in Sec.~\ref{sec:cosmic-strings} we can make the simple assertion that $\upDelta t\approx1/\nu_\mathrm{o}$~\cite{Regimbau:2011bm}.

We can write the physical volume element in our perturbed FLRW metric as
    \begin{equation}
        \dd[3]{V}=\dd[2]{\sigma_\mathrm{o}}\dd{\eta}a^3r^2\qty(1+\Psi+\Pi+\vu*e_\mathrm{o}\vdot\vb*v),
    \end{equation}
    where $r$ is the comoving distance measure, written in terms of the conformal time as
    \begin{equation}
        r\qty(\eta)\equiv\int_\eta^{\eta_\mathrm{o}}\dd{\eta}\qty(1+2\Psi).
    \end{equation}
Integrating over solid angle averages out the cosmological perturbations, and hence the cutoff time $\eta_*$ is found by solving the integral equation
    \begin{equation}
        \frac{4\uppi}{\nu_\mathrm{o}}\int\dd{\vb*\zeta}\int_{\eta_*}^{\eta_\mathrm{o}}\dd{\eta}a^3\qty(\eta_\mathrm{o}-\eta)^2f_\mathrm{o}\bar{n}R=1.
    \end{equation}
We therefore modify the conformal time integral in our previously found linear-order expression Eq.~\eqref{eq:omega-gw} and get
    \begin{align}
    \begin{split}
        \label{eq:omega-gw-final}
        \Omega_\mathrm{gw}\qty(\nu_\mathrm{o},\vu*e_\mathrm{o})=&\frac{\uppi\nu_\mathrm{o}^3}{3H_\mathrm{o}^2}\int_0^{\eta_*}\dd{\eta}a^2\int\dd{\vb*\zeta}\bar{n}R\bigg[1+\delta_n-\Psi_\mathrm{o}-\Pi_\mathrm{o}\\
        &+2\qty(\Psi+\Pi)+\vu*e_\mathrm{o}\vdot\vb*v_\mathrm{o}+2\int_\eta^{\eta_\mathrm{o}}\dd{\eta'}\pdv{\Psi}{\eta'}\bigg]\\
        &\times\int_{S^2}\dd[2]{\sigma_\mathrm{s}}r\mathrlap{^2}_\mathrm{s}\tilde{h}^2.
    \end{split}
    \end{align}
This expression Eq.~\eqref{eq:omega-gw-final} is the main result of our analysis; it can be used for any astrophysical or cosmological source of anisotropies in the stochastic background of gravitational waves.

\subsection{Characterizing the anisotropies}
\label{sec:formalism-sgwb-decomposition}
We will initially focus on the anisotropy due to the source density contrast $\delta_n$, and therefore neglect most of the cosmological perturbations.
The only other term we include is the peculiar motion of the observer $\vb*v_\mathrm{o}$, as this introduces a ``kinematic dipole" that interferes with the anisotropy statistics.
In the case of the cosmic microwave background (CMB), this dipole is roughly 100 times greater than the ``true" cosmological fluctuations we are interested in, so it is usually subtracted from the raw data before calculating any statistics. We will do the same for the SGWB.

There are two possible approaches to this: either measure the observed kinematic dipole of the SGWB directly at each frequency and subtract it, or use CMB data to measure the direction of the dipole, and use the formalism discussed above to generate a theoretical prediction for its magnitude.
Since SGWB measurements are likely to be much less precise than CMB measurements in both overall magnitude and angular resolution for the foreseeable future, the latter seems  to us the best approach.

Thus, setting $\Psi=\Pi=0$ everywhere and $\vb*v=\vb*0$ everywhere except at the observer, we have
    \begin{align}
    \begin{split}
        \Omega_\mathrm{gw}\qty(\nu_\mathrm{o},\vu*e_\mathrm{o})=\frac{\uppi\nu_\mathrm{o}^3}{3H_\mathrm{o}^2}&\int_0^{\eta_*}\dd{\eta}a^2\int\dd{\vb*\zeta}\bar{n}R\\
        &\times\qty(1+\delta_n+\vu*e_\mathrm{o}\vdot\vb*v_\mathrm{o})\int_{S^2}\dd[2]{\sigma_\mathrm{s}}r\mathrlap{^2}_\mathrm{s}\tilde{h}^2,
    \end{split}
    \end{align}
    with the emitted frequency given by
    \begin{equation}
        \label{eq:doppler-shift}
        \nu_\mathrm{s}=\frac{\nu_\mathrm{o}}{a}\qty(1-\vu*e_\mathrm{o}\vdot\vb*v_\mathrm{o}).
    \end{equation}
We thus see that the observer's peculiar motion causes a Doppler shift in the observed frequencies for each source, which will vary in importance depending on the cosmological redshifts of the sources.
This means that the magnitude of kinematic dipole will depend on the waveform $\tilde{h}$ and distance of every source that contributes to the SGWB, making the required calculation more complicated than that for the CMB dipole.
We sketch here how to calculate the size of the dipole, with a more concrete treatment for the cosmic string case given in Sec.~\ref{eq:cs-sgwb-decomposition}.

As we are working only to linear order, we define
    \begin{equation}
        x\qty(\vu*e_\mathrm{o})\equiv1+\vu*e_\mathrm{o}\vdot\vb*v_\mathrm{o}
    \end{equation}
    and express all modifications due to the kinematic dipole as powers of $x$.
This depends only on $\vu*e_\mathrm{o}$, and is therefore unaffected by the integrals over $\vb*\zeta$ and $\eta$.
With reference to Eq.~\eqref{eq:monopole-density-contrast}, we see that the averaged isotropic background value (monopole) is given by
    \begin{equation}
        \bar{\Omega}_\mathrm{gw}\qty(\nu_\mathrm{o})\equiv\frac{1}{4\uppi}\int_{S^2}\dd[2]{\sigma_\mathrm{o}}\Omega_\mathrm{gw}\qty(\nu_\mathrm{o},\vu*e_\mathrm{o})=\left.\Omega_\mathrm{gw}\right|_{x=1,\delta_n=0}
    \end{equation}
    with the anisotropies described by the SGWB energy density contrast,
    \begin{equation}
        \delta_\mathrm{gw}(\nu_\mathrm{o},\vu*e_\mathrm{o})\equiv\frac{\Omega_\mathrm{gw}-\bar{\Omega}_\mathrm{gw}}{\bar{\Omega}_\mathrm{gw}}.
    \end{equation}
The quantity we are interested in is the density contrast due to the source distribution alone, with the kinematic dipole subtracted.
This is defined as
    \begin{equation}
        \delta_\mathrm{gw}^\qty(\mathrm{s})(\nu_\mathrm{o},\vu*e_\mathrm{o})\equiv\left.\delta_\mathrm{gw}\right|_{x=1}=\frac{\left.\Omega_\mathrm{gw}\right|_{x=1}-\bar{\Omega}_\mathrm{gw}}{\bar{\Omega}_\mathrm{gw}}
    \end{equation}
    where ``s" stands for ``source".
We can compute the linear-order correction due to the kinematic dipole with a Taylor expansion around $x=1$,
    \begin{align*}
        \Omega_\mathrm{gw}&=\left.\Omega_\mathrm{gw}\right|_{x=1}+\vu*e_\mathrm{o}\vdot\vb*v_\mathrm{o}\left.\pdv{\Omega_\mathrm{gw}}{x}\right|_{x=1}\\
        &=\bar{\Omega}_\mathrm{gw}\qty(1+\delta_\mathrm{gw}^\qty(\mathrm{s}))+\vu*e_\mathrm{o}\vdot\vb*v_\mathrm{o}\left.\pdv{\Omega_\mathrm{gw}}{x}\right|_{x=1,\delta_n=0},
    \end{align*}
    where the latter equality holds because $\vu*e_\mathrm{o}\vdot\vb*v_\mathrm{o}\delta_n$ is second order.
We therefore find
    \begin{align}
    \begin{split}
        \label{eq:dipole-factor}
        \delta_\mathrm{gw}&=\delta_\mathrm{gw}^\qty(\mathrm{s})+\mathcal{D}\,\vu*e_\mathrm{o}\vdot\vu*v_\mathrm{o},\\
        \mathcal{D}&\equiv v_\mathrm{o}\bar{\Omega}_\mathrm{gw}^{-1}\left.\pdv{\Omega_\mathrm{gw}}{x}\right|_{x=1,\delta_n=0},
    \end{split}
    \end{align}
    where $v_\mathrm{o}\equiv\qty|\vb*v_\mathrm{o}|$, $\vu*v_\mathrm{o}\equiv\vb*v_\mathrm{o}/v_\mathrm{o}$, and $\mathcal{D}\qty(\nu_\mathrm{o})$ is a frequency-dependent coefficient describing the size of the kinematic dipole, which depends on the GW waveforms and spatial distribution of the sources.
Note that this approach is only valid if $\delta^{(\mathrm{s})}_\mathrm{gw}\gg v_\mathrm{o}^2$; otherwise we must go beyond the linear expansion.

Now we are able to study $\delta^{(\mathrm{s})}_\mathrm{gw}$, either directly or in terms of its statistics.
One particularly useful statistical descriptor is the two-point correlation function (2PCF), defined as the second moment of the density contrast,
    \begin{equation}
        C_\mathrm{gw}\qty(\theta_\mathrm{o},\nu_\mathrm{o})\equiv\ev{\delta_\mathrm{gw}^\qty(\mathrm{s})\qty(\nu_\mathrm{o},\vu*e_\mathrm{o})\delta_\mathrm{gw}^\qty(\mathrm{s})(\nu_\mathrm{o},\vu*e\mathrlap{'}_\mathrm{o})},
    \end{equation}
    where $\theta_\mathrm{o}\equiv\cos^{-1}(\vu*e_\mathrm{o}\vdot\vu*e\mathrlap{'}_\mathrm{o})$, and the angle brackets denote an averaging over all pairs of directions $\vu*e_\mathrm{o}$, $\vu*e\mathrlap{'}_\mathrm{o}$ whose angle of separation is $\theta_\mathrm{o}$.
The first moment (i.e. mean) vanishes by definition, and if the background is a GRF (as discussed in \ref{sec:gaussian-non-gaussian}) then all higher moments either vanish or are expressed in terms of the second moment by Wick's theorem.
The 2PCF therefore uniquely characterizes the anisotropies in the Gaussian part of the background.
It is common practice (particularly in the CMB literature) to perform a multipole expansion of the 2PCF,
    \begin{equation}
        \label{eq:multipole-expansion}
        C_\mathrm{gw}\qty(\theta_\mathrm{o},\nu_\mathrm{o})=\sum_{\ell=0}^\infty\frac{2\ell+1}{4\uppi}C_\ell\qty(\nu_\mathrm{o})P_\ell(\cos\theta_\mathrm{o}),
    \end{equation}
    where $P_\ell\qty(x)$ denotes the $\ell\mathrm{th}$ Legendre polynomial.
The anisotropies are then described in terms of the $C_\ell$ components, which are given by
    \begin{equation}
        C_\ell\qty(\nu_\mathrm{o})\equiv2\uppi\int_{-1}^{+1}\dd{\qty(\cos\theta_\mathrm{o})}C_\mathrm{gw}\qty(\theta_\mathrm{o},\nu_\mathrm{o})P_\ell\qty(\cos\theta_\mathrm{o}).
    \end{equation}
The quantity $\ell\qty(\ell+1)C_\ell/2\uppi$ is roughly the contribution to the variance of $\delta^{(\mathrm{s})}_\mathrm{gw}$ per logarithmic bin in $\ell$, as can be seen by considering
    \begin{equation*}
        \mathrm{var}\qty(\delta^{(\mathrm{s})}_\mathrm{gw})=\sum_\ell\frac{2\ell+1}{4\uppi}C_\ell\approx\int\dd{\qty(\ln\ell)}\frac{\ell\qty(\ell+1)}{2\uppi}C_\ell.
    \end{equation*}

Defined in this way, the 2PCF excludes the kinematic dipole.
The effects of including this on the $C_\ell$ components are described in the Appendix.

\subsection{Estimating the 2PCF from observations}
\label{sec:observations}
The decomposition of the 2PCF described above is not the only way of describing the SGWB anisotropies.
Another convenient tool is the spherical harmonic decomposition of $\Omega_\mathrm{gw}$ itself,
    \begin{equation}
        \Omega_\mathrm{gw}(\nu_\mathrm{o},\vu*e_\mathrm{o})=\sum_{\ell=0}^\infty\sum_{m=-\ell}^{+\ell}\Omega_{\ell m}(\nu_\mathrm{o})Y_{\ell m}(\vu*e_\mathrm{o}),
    \end{equation}
    where $Y_{\ell m}\qty(\vu*e_\text{o})$ are the Laplace spherical harmonics, and
    \begin{equation}
        \Omega_{\ell m}(\nu_\mathrm{o})\equiv\int_{S^2}\dd[2]{\sigma_\mathrm{o}}\Omega_\mathrm{gw}(\nu_\mathrm{o},\vu*e_\mathrm{o})Y\mathrlap{^*}_{\ell m}(\vu*e_\mathrm{o}).
    \end{equation}
We can perform the same decomposition for $\delta_\mathrm{gw}$,
    \begin{align}
    \begin{split}
        \delta_\mathrm{gw}(\nu_\mathrm{o},\vu*e_\mathrm{o})&=\sum_{\ell=0}^\infty\sum_{m=-\ell}^{+\ell}\omega_{\ell m}(\nu_\mathrm{o})Y_{\ell m}(\vu*e_\mathrm{o}),\\
        \omega_{\ell m}(\nu_\mathrm{o})&\equiv\int_{S^2}\dd[2]{\sigma_\mathrm{o}}\delta_\mathrm{gw}(\nu_\mathrm{o},\vu*e_\mathrm{o})Y\mathrlap{^*}_{\ell m}(\vu*e_\mathrm{o}),
    \end{split}
    \end{align}
    with the $\omega_{\ell m}$ components given in terms of the $\Omega_{\ell m}$'s by
    \begin{equation}
        \label{eq:omega-Omega-lm}
        \omega_{\ell m}=\bar{\Omega}_\mathrm{gw}^{-1}\Omega_{\ell m}-\sqrt{4\uppi}\delta_{\ell0}\delta_{m0}.
    \end{equation}
Here we have used the orthogonality condition for the spherical harmonics
    \begin{equation}
        \int_{S^2}\dd[2]{\sigma_\mathrm{o}}Y_{\ell m}\qty(\vu*e_\mathrm{o})Y\mathrlap{^*}_{\ell'm'}\qty(\vu*e_\mathrm{o})=\delta_{\ell\ell'}\delta_{mm'},
    \end{equation}
    and the fact that $Y_{00}=1/\sqrt{4\uppi}$.

Since we are interested in the $C_\ell$'s of the source anisotropies $\delta^{(\mathrm{s})}_\mathrm{gw}$, we want to remove the kinematic dipole from Eq.~\eqref{eq:omega-Omega-lm}.
Doing so inevitably involves a particular choice of co\"ordinates $\vu*e_\mathrm{o}=\qty(\theta_\mathrm{o},\phi_\mathrm{o})$.
For simplicity, we take the direction of the kinematic dipole $\vu*v_\mathrm{o}$ as the $\theta_\mathrm{o}=0$ direction, so that
    \begin{equation}
        \delta^{(\mathrm{s})}_\mathrm{gw}=\delta_\mathrm{gw}-\mathcal{D}\cos\theta_\mathrm{o}.
    \end{equation}
The dipole is then proportional to $Y_{10}=\sqrt{3/4\uppi}\cos\theta_\mathrm{o}$, so performing the decomposition,
    \begin{align}
    \begin{split}
        \delta^{(\mathrm{s})}_\mathrm{gw}(\nu_\mathrm{o},\vu*e_\mathrm{o})&=\sum_{\ell=0}^\infty\sum_{m=-\ell}^{+\ell}\omega^{(\mathrm{s})}_{\ell m}(\nu_\mathrm{o})Y_{\ell m}(\vu*e_\mathrm{o}),\\
        \omega^{(\mathrm{s})}_{\ell m}(\nu_\mathrm{o})&\equiv\int_{S^2}\dd[2]{\sigma_\mathrm{o}}\delta^{(\mathrm{s})}_\mathrm{gw}(\nu_\mathrm{o},\vu*e_\mathrm{o})Y\mathrlap{^*}_{\ell m}(\vu*e_\mathrm{o}),
    \end{split}
    \end{align}
    we see that Eq.~\eqref{eq:omega-Omega-lm} becomes
    \begin{equation}
        \label{eq:omega-Omega-lm-dipole}
        \omega^{(\mathrm{s})}_{\ell m}=\bar{\Omega}_\mathrm{gw}^{-1}\Omega_{\ell m}-\sqrt{4\uppi}\delta_{\ell0}\delta_{m0}-\sqrt{\frac{4\uppi}{3}}\mathcal{D}\delta_{\ell1}\delta_{m0}.
    \end{equation}

The relationship between these spherical harmonic decompositions and the $C_\ell$ components can be found by writing
    \begin{align*}
        C_\mathrm{gw}&\equiv\ev{\delta^{(\mathrm{s})}_\mathrm{gw}(\vu*e_\text{o})\delta^{(\mathrm{s})}_\mathrm{gw}(\vu*e\mathrlap{'}_\text{o})}\\
        &=\sum_{\ell=0}^\infty\sum_{\ell'=0}^\infty\sum_{m=-\ell}^{+\ell}\sum_{m'=-\ell'}^{+\ell'}\ev{\omega^{(\mathrm{s})}_{\ell m}\omega^{(\mathrm{s})*}_{\ell'm'}}Y_{\ell m}(\vu*e_\text{o})Y\mathrlap{^*}_{\ell'm'}(\vu*e\mathrlap{'}_\text{o})\\
        &=\sum_{\ell=0}^\infty\frac{2\ell+1}{4\uppi}C_\ell P_\ell(\vu*e_\text{o}\vdot\vu*e\mathrlap{'}_\text{o}).
    \end{align*}
We require the RHS above to be invariant under rotations of the sphere, which implies that $\ev{\omega^{(\mathrm{s})}_{\ell m}\omega^{(\mathrm{s})*}_{\ell'm'}}$ is proportional to $\delta_{\ell\ell'}\delta_{mm'}$.
Using the addition theorem for spherical harmonics,
    \begin{equation}
        \sum_{m=-\ell}^{+\ell}Y_{\ell m}(\vu*e_\text{o})Y\mathrlap{^*}_{\ell m}(\vu*e\mathrlap{'}_\text{o})=\frac{2\ell+1}{4\uppi}P_\ell(\vu*e_\text{o}\vdot\vu*e\mathrlap{'}_\text{o}),
    \end{equation}
    we therefore see that
    \begin{equation}
        \ev{\omega^{(\mathrm{s})}_{\ell m}\omega^{(\mathrm{s})*}_{\ell'm'}}=C_\ell\delta_{\ell\ell'}\delta_{mm'},
    \end{equation}
    and thus
    \begin{equation}
        C_\ell=\frac{1}{2\ell+1}\sum_{m=-\ell}^{+\ell}\ev{\omega^{(\mathrm{s})}_{\ell m}\omega^{(\mathrm{s})*}_{\ell m}}
    \end{equation}
    which directly relates the $C_\ell$'s to the $\omega^{(\mathrm{s})}_{\ell m}$'s~\footnote{This result, and indeed much of this section, is directly analogous to the corresponding CMB result. For a detailed treatment of these issues in the case of the CMB, we refer the reader to Ref.~\cite{Durrer:2008eom}.}.
Note that the angle brackets here indicate an ensemble average over random realizations of the $\Omega_\mathrm{gw}$ field.

This expression shows that the $\omega^{(\mathrm{s})}_{\ell m}$ components contain more information about each random realization of the SGWB than the $C_\ell$'s do.
There is an averaging process (the angle brackets) that takes us from the $\omega^{(\mathrm{s})}_{\ell m}$'s to the $C_\ell$'s (or, equivalently, from $\Omega_\mathrm{gw}$ to $C_\mathrm{gw}$), so there must be many possible configurations of the field $\Omega_\mathrm{gw}$ that all correspond to the same $C_\ell$'s but give different $\omega^{(\mathrm{s})}_{\ell m}$'s.
This means that we cannot invert the above equation and reconstruct $\Omega_\mathrm{gw}$ in terms of the $C_\ell$'s alone.

With a view towards future observational work, we can relate the $\omega^{(\mathrm{s})}_{\ell m}$ and $C_\ell$ components to the GW strain $h_{ij}$ measured by the observer.
This is given by
    \begin{equation}
        h_{ij}\qty(t_\text{o})=\sum_{A=+,\times}\int_{S^2}\dd[2]{\sigma_\mathrm{o}}\int_{-\infty}^{+\infty}\dd{\nu_\mathrm{o}}\tilde{h}_Ae^A_{ij}\mathrm{e}^{2\uppi\mathrm{i}\nu_\mathrm{o}t_\mathrm{o}},
    \end{equation}
    where $e^+_{ij}$, $e^\times_{ij}$ are polarization tensors and $\tilde{h}_+$, $\tilde{h}_\times$ are the Fourier components of the background \cite{Maggiore:1999vm}.
The signal is often characterized by the quadratic expectation value of these Fourier components.
For a SGWB that is unpolarized, Gaussian, and stationary (but still anisotropic), these expectation values can be written as \cite{Romano:2016dpx}
    \begin{align}
    \begin{split}
        &\ev{\tilde{h}_A(\nu_\mathrm{o},\vu*e_\mathrm{o})\tilde{h}_{A'}(\nu\mathrlap{'}_\mathrm{o},\vu*e\mathrlap{'}_\mathrm{o})}\\
        &\qquad=\frac{1}{4}\mathcal{P}(\nu_\mathrm{o},\vu*e_\mathrm{o})\delta(\nu_\mathrm{o}-\nu\mathrlap{'}_\mathrm{o})\delta_{AA'}\delta^{(2)}(\vu*e_\mathrm{o}-\vu*e\mathrlap{'}_\mathrm{o}),
    \end{split}
    \end{align}
    where $\mathcal{P}$ is the power spectrum.
This can be written in terms of the density parameter as
    \begin{equation}
        \label{eq:power-spectrum-def}
        \mathcal{P}(\nu_\mathrm{o},\vu*e_\mathrm{o})=\frac{3H_\mathrm{o}^2}{2\uppi^2\nu_\mathrm{o}^3}\Omega_\mathrm{gw}(\nu_\mathrm{o},\vu*e_\mathrm{o}).
    \end{equation}
The corresponding isotropic quantity is the power spectral density (PSD),
    \begin{equation}
        S_h(\nu_\mathrm{o})\equiv\int_{S^2}\dd[2]{\sigma_\mathrm{o}}\mathcal{P}(\nu_\mathrm{o},\vu*e_\mathrm{o})=\frac{6H_\mathrm{o}^2}{\uppi\nu_\mathrm{o}^3}\bar{\Omega}_\mathrm{gw}.
    \end{equation}
(Note that this differs from the usual expression by a factor of $4\uppi$, due to our definition of the monopole $\bar{\Omega}_\mathrm{gw}$.)

The power spectrum can itself be decomposed in spherical harmonics,
    \begin{align}
    \begin{split}
        \mathcal{P}(\nu_\mathrm{o},\vu*e_\mathrm{o})&=\sum_{\ell=0}^\infty\sum_{m=-\ell}^{+\ell}\mathcal{P}_{\ell m}(\nu_\mathrm{o})Y_{\ell m}(\vu*e_\mathrm{o}),\\
        \mathcal{P}_{\ell m}(\nu_\mathrm{o})&\equiv\int_{S^2}\dd[2]{\sigma_\mathrm{o}}\mathcal{P}(\nu_\mathrm{o},\vu*e_\mathrm{o})Y\mathrlap{^*}_{\ell m}(\vu*e_\mathrm{o}).
    \end{split}
    \end{align}
Observational efforts to detect an anisotropic background are commonly phrased in terms of these $\mathcal{P}_{\ell m}$ components \cite{Romano:2016dpx}, so it is valuable to relate these to the $C_\ell$'s computed in this work.
We relate them first to the $\omega_{\ell m}$'s using Eqs.~\eqref{eq:omega-Omega-lm-dipole} and \eqref{eq:power-spectrum-def} to give
    \begin{align}
    \begin{split}
        \mathcal{P}_{\ell m}&=\frac{3H_\mathrm{o}^2}{2\uppi^2\nu_\mathrm{o}^3}\Omega_{\ell m}\\
        &=\frac{S_h}{4\uppi}\qty[\sqrt{4\uppi}\delta_{\ell0}\delta_{m0}+\sqrt{\frac{4\uppi}{3}}\mathcal{D}\delta_{\ell1}\delta_{m0}+\omega^{(\mathrm{s})}_{\ell m}].
    \end{split}
    \end{align}
We then use the above to relate the $\mathcal{P}_{\ell m}$'s to the $C_\ell$'s,
    \begin{align}
    \begin{split}
        C_\ell=\frac{1}{2\ell+1}&\sum_{m=-\ell}^{+\ell}\ev{\omega^{(\mathrm{s})}_{\ell m}\omega^{(\mathrm{s})*}_{\ell m}}\\
        =\frac{16\uppi^2}{2\ell+1}&\qty[\sum_{m=-\ell}^{+\ell}\frac{\ev{\mathcal{P}_{\ell m}\mathcal{P}\mathrlap{^*}_{\ell m}}}{S_h^2}]+\qty[4\uppi-16\uppi^{3/2}\frac{\ev{\mathcal{P}_{00}}}{S_h}]\delta_{\ell0}\\
        &+\qty[\frac{4\uppi}{9}\mathcal{D}^2-16\qty(\frac{\uppi}{3})^{3/2}\mathcal{D}\frac{\ev{\mathcal{P}_{10}}}{S_h}]\delta_{\ell1}.
    \end{split}
    \end{align}
This slightly cumbersome expression is due to the fact that we are expressing the $C_\ell$'s for the 2PCF of the density \emph{contrast} $\delta^{(\mathrm{s})}_\mathrm{gw}$ in terms of the power spectrum of the density itself, $\Omega_\text{gw}$.
Normalizing the density with respect to its average isotropic value causes the $\mathcal{P}_{\ell m}$'s to be normalized relative to the PSD $S_h$, while removing the monopole and kinematic dipole gives rise to extra terms for the $\ell=0$ and $\ell=1$ modes, respectively.

We cannot perform the ensemble average over the $\mathcal{P}_{\ell m}$'s implied by the angle brackets here, as we only have one realization of the SGWB.
However, the above expression gives an obvious choice of an estimator for each $C_\ell$, where we use the measured value of each $\mathcal{P}_{\ell m}$ for our particular realization of the SGWB in lieu of an ensemble average:
\begin{widetext}
    \begin{equation}
        \label{eq:c-l-estimator}
        \hat{C}_\ell=\frac{16\uppi^2}{\qty(2\ell+1)S_h^2}
        \begin{cases}
            \qty(\mathcal{P}_{00}-\frac{S_h}{\sqrt{4\uppi}})^2,&\ell=0,\\
            \qty(\mathcal{P}_{10}-\frac{\mathcal{D}S_h}{\sqrt{12\uppi}})^2+\qty|\mathcal{P}_{1-1}|^2+\qty|\mathcal{P}_{11}|^2,&\ell=1,\\
            \sum_{m=-\ell}^{+\ell}\mathcal{P}_{\ell m}\mathcal{P}\mathrlap{^*}_{\ell m},&\ell>1.
        \end{cases}
    \end{equation}
\end{widetext}
We see that this is unbiased (i.e. the mean of the estimator is equal to the estimated quantity, $\ev*{\hat{C}_\ell}=C_\ell$).
By analogy with the CMB, we call the variance of this estimator the \emph{cosmic variance}.
This is the error associated with the fact that we only have one random realization of the SGWB.
For a Gaussian background, the $\omega^{(\mathrm{s})}_{\ell m}$'s are all zero-mean Gaussian fields, so Wick's theorem gives
    \begin{align*}
        \ev{\omega^{(\mathrm{s})}_{\ell m}\omega^{(\mathrm{s})*}_{\ell m}\omega^{(\mathrm{s})}_{\ell m'}\omega^{(\mathrm{s})*}_{\ell m'}}&=\ev{\omega^{(\mathrm{s})}_{\ell m}\omega^{(\mathrm{s})*}_{\ell m}}\ev{\omega^{(\mathrm{s})}_{\ell m'}\omega^{(\mathrm{s})*}_{\ell m'}}\\
        &\quad+\ev{\omega^{(\mathrm{s})}_{\ell m}\omega^{(\mathrm{s})}_{\ell m'}}\ev{\omega^{(\mathrm{s})*}_{\ell m}\omega^{(\mathrm{s})*}_{\ell m'}}\\
        &\quad+\ev{\omega^{(\mathrm{s})}_{\ell m}\omega^{(\mathrm{s})*}_{\ell m'}}\ev{\omega^{(\mathrm{s})}_{\ell m'}\omega^{(\mathrm{s})*}_{\ell m}}.
    \end{align*}
The cosmic variance is then easy to evaluate
    \begin{align}
    \begin{split}
        \label{eq:cosmic-var}
        &\mathrm{var}(\hat{C}_\ell)\equiv\ev*{\hat{C}_\ell\hat{C}_\ell}-\ev*{\hat{C}_\ell}^2\\
        &=\frac{1}{(2\ell+1)^2}\sum_{m=-\ell}^{+\ell}\sum_{m'=-\ell}^{+\ell}\ev{\omega^{(\mathrm{s})}_{\ell m}\omega^{(\mathrm{s})*}_{\ell m}\omega^{(\mathrm{s})}_{\ell m'}\omega^{(\mathrm{s})*}_{\ell m'}}\\
        &\qquad\qquad\qquad\qquad\qquad\qquad-\ev{\omega^{(\mathrm{s})}_{\ell m}\omega^{(\mathrm{s})*}_{\ell m}}\ev{\omega^{(\mathrm{s})}_{\ell m'}\omega^{(\mathrm{s})*}_{\ell m'}}\\
        &=\frac{1}{(2\ell+1)^2}\sum_{m=-\ell}^{+\ell}\sum_{m'=-\ell}^{+\ell}\ev{\omega^{(\mathrm{s})}_{\ell m}\omega^{(\mathrm{s})}_{\ell m'}}\ev{\omega^{(\mathrm{s})*}_{\ell m}\omega^{(\mathrm{s})*}_{\ell m'}}\\
        &\qquad\qquad\qquad\qquad\qquad\qquad+\ev{\omega^{(\mathrm{s})}_{\ell m}\omega^{(\mathrm{s})*}_{\ell m'}}\ev{\omega^{(\mathrm{s})}_{\ell m'}\omega^{(\mathrm{s})*}_{\ell m}}\\
        &=\frac{2}{(2\ell+1)^2}\sum_{m=-\ell}^{+\ell}\ev{\omega^{(\mathrm{s})}_{\ell m}\omega^{(\mathrm{s})*}_{\ell m}}\ev{\omega^{(\mathrm{s})}_{\ell m'}\omega^{(\mathrm{s})*}_{\ell m'}}\\
        &=\frac{2}{2\ell+1}C_\ell^2.
    \end{split}
    \end{align}
This is exactly the same as the equivalent result for the CMB temperature anisotropies, although the $C_\ell$'s themselves are of course different.

Thus, Eq.~\eqref{eq:c-l-estimator} tells us how best to reconstruct the $C_\ell$ components of the SGWB from the observed values of the $\mathcal{P}_{\ell m}$, with the cosmic variance given by Eq.~\eqref{eq:cosmic-var}.
Note that this is only valid for a Gaussian background, so it is important that the condition in Eq.~\eqref{eq:grf-condition} is satisfied.
As before, the $C_\ell$ components discussed in this section do not include the kinematic dipole, but this can be included using the results in the Appendix.

\subsection{A note on previous works}
We draw the reader's attention to two recent articles that are of relevance.
The first, Ref.~\cite{Cusin:2017fwz}, introduces much of the relevant formalism, and offers a thorough derivation of Eqs.~\eqref{eq:cusin-et-al-main-result} and \eqref{eq:source-frequency}, which served as our starting point.
We emphasize that the additional formalism introduced in Sec.~\ref{sec:relating-strain-luminosity}--\ref{sec:observations} goes beyond what was done in Ref.~\cite{Cusin:2017fwz}.
The second, Ref.~\cite{Cusin:2017mjm}, expresses the results of Ref.~\cite{Cusin:2017fwz} in terms of the power spectrum of a quantity $\dd[2]{\mathcal{Q}_A}$, which is related to the emitted strain.
However, this quantity is defined as a combination of several distinct physical variables, and is an inconvenient choice of description for the anisotropies.
The 2PCF that we have used above is an alternative (and, we believe, clearer and more practical) means of characterizing the anisotropies, which lends itself better to concrete calculations and comparisons with observations.

\section{Cosmic Strings}
\label{sec:cosmic-strings}
Cosmic strings are one-dimensional topological defects formed in the early Universe as a result of a phase transition, followed by a spontaneous symmetry breaking characterized by a vacuum manifold with non-contractible closed curves~\cite{Kibble:1976sj}. These linear defects are expected to be generically produced in the context of grand unified theories~\cite{Jeannerot:2003qv}.
Subhorizon cosmic strings (so-called ``loops") oscillate periodically in time, emitting GWs as they do so; superhorizon strings (so-called ``infinite strings") also emit GWs, since they are not straight and have small-scale structure as the result of string intercommutations~\cite{Sakellariadou:1990ne}.

In what follows, we use the formalism presented in the previous sections to calculate the expected SGWB due to GW bursts from a network of cosmic string loops.
The waveforms of these bursts are given by expansions in $1/r_\mathrm{s}$, so that in the transverse traceless gauge they read~\cite{Damour:2001bk}
    \begin{equation}
        h_{ij}\qty(t_\mathrm{s},\vb*x_\mathrm{s})=\frac{\kappa_{ij}\qty(t_\mathrm{s}-r_\mathrm{s},\vu*e_\mathrm{s})}{r_\mathrm{s}}+\order{\frac{1}{r\mathrlap{^2}_\mathrm{s}}},
    \end{equation}
    with $r_\mathrm{s}\equiv\qty|\vb*x_\mathrm{s}|$ and $\vu*e_\mathrm{s}\equiv\vb*x_\mathrm{s}/r_\mathrm{s}$.
In the local wave zone, one can consider $r_\mathrm{s}$ much greater than the size of the source, and thus neglect subleading terms in this expansion.
Therefore,
    \begin{equation*}
        \tilde{h}\qty(\nu_\mathrm{s})\approx\frac{\kappa\qty(\nu_\mathrm{s},\vu*e_\mathrm{s})}{\nu_\mathrm{s}r_\mathrm{s}},
    \end{equation*}
    where we follow Ref.~\cite{Damour:2001bk} in defining $\kappa\qty(\nu_\mathrm{s},\vu*e_\mathrm{s})$ \emph{not} as the Fourier transform of $\kappa\qty(t_\mathrm{s},\vu*e_\mathrm{s})$, but as
    \begin{equation}
        \kappa\qty(\nu_\mathrm{s},\vu*e_\mathrm{s})\equiv\nu_\mathrm{s}\tilde{\kappa}\qty(\nu_\mathrm{s},\vu*e_\mathrm{s})=\nu_\mathrm{s}\int_{-\infty}^{+\infty}\dd{t_\mathrm{s}}\mathrm{e}^{2\uppi\mathrm{i}\nu_\mathrm{s}t_\mathrm{s}}\kappa\qty(t_\mathrm{s},\vu*e_\mathrm{s}),
    \end{equation}
    giving it the same units as $\kappa\qty(t_\mathrm{s},\vu*e_\mathrm{s})$.

The main simplification in the string loop case is that instead of having a multitude of parameters $\vb*\zeta$ describing the sources, there is just one: the fundamental loop length $l$.
All we need to know is $n\qty(\eta,\vu*e_\mathrm{o},l)$, the loop number density distribution with respect to $l$.
In an expanding Universe, superhorizon sized loops reach a scaling solution with respect to cosmic time $t$ in which the relative length $l/t$ is constant.
It is therefore convenient to define the dimensionless quantities~\cite{Lorenz:2010sm}
    \begin{equation}
        \gamma\equiv\frac{l}{t},\quad\mathcal{F}\qty(t,\vu*e_\mathrm{o},\gamma)\equiv t^4n\qty(t,\vu*e_\mathrm{o},l),\quad\bar{\mathcal{F}}\qty(\gamma)\equiv t^4\bar{n}\qty(t,l),
    \end{equation}
    where $\bar{\mathcal{F}}$ is the homogeneous scaling solution, which is constant in time.
We can therefore simplify the distribution in $l$ at time $t$ to a distribution in $\gamma$, keeping in mind to integrate over $\dd{l}=t\dd{\gamma}$ and not just $\dd{\gamma}$.
Being interested in subhorizon loops, we set $\gamma\in\qty[0,\gamma_*]$, where
    \begin{equation}
        \gamma_*(t)\equiv\frac{a}{t}\int_0^t\frac{\dd{t'}}{a\qty(t')}
    \end{equation}
    is the relative physical horizon size.
Note that
    \begin{equation}
        \delta_\mathcal{F}\qty(t,\vu*e_\mathrm{o},\gamma)\equiv\frac{\mathcal{F}-\bar{\mathcal{F}}}{\bar{\mathcal{F}}}=\frac{n-\bar{n}}{\bar{n}}=\delta_n.
    \end{equation}

We will consider several different types of GW burst events, which we label with a subscript $i$.
For each loop, we write the rate of bursts of type $i$ as
    \begin{equation}
        R_i=\frac{N_i}{T}=\frac{2N_i}{l},
    \end{equation}
    where $T=l/2$ is the loop oscillation period (with corresponding fundamental frequency $2/l$) and $N_i$ is the number of bursts of type $i$ per oscillation. Such an oscillating loop emitting GWs decays in a lifetime $l/\gamma_{\mathrm{d}}$, with $\gamma_{\mathrm{d}}$ the \emph{gravitational decay scale}.

Using the above, and integrating over $\dd{t}=a\dd{\eta}$ rather than $\dd{\eta}$, Eq.~\eqref{eq:omega-gw-final} in the case of cosmic strings becomes
    \begin{align}
    \begin{split}
        \Omega_\mathrm{gw}=\frac{2\uppi\nu_\mathrm{o}}{3H_\mathrm{o}^2}\int_0^{t_*}\frac{\dd{t}}{t^4}a^3&\int_0^{\gamma_*}\frac{\dd{\gamma}}{\gamma}\bar{\mathcal{F}}\qty(1+\delta_\mathcal{F}+3\vu*e_\mathrm{o}\vdot\vb*v_\mathrm{o})\\
        &\times\qty(\sum_iN_i\int_{S^2}\dd[2]{\sigma_\mathrm{s}}\kappa_i^2),
    \end{split}
    \end{align}
    where $t_*$ is defined by
    \begin{equation}
        \Lambda\qty(\nu_\text{o},t_*)=\frac{8\uppi}{\nu_\mathrm{o}}\int_{t_*}^{t_\mathrm{o}}\frac{\dd{t}}{t^4}a^2r^2\int_0^{\gamma_*}\frac{\dd{\gamma}}{\gamma}\bar{\mathcal{F}}\sum_iN_i\bar{f}_{\mathrm{o},i}=1,
    \end{equation}
    and the comoving distance is
    \begin{equation}
        r\qty(t)=\int_t^{t_\mathrm{o}}\frac{\dd{t'}}{a\qty(t')}.
    \end{equation}

\subsection{Cusps, kinks, and kink-kink collisions}
\label{sec:cusps-kinks-collisions}
Usually, two types of bursts are identified: those associated with points on the string briefly reaching the speed of light (called ``cusps"), and those associated with discontinuities in the string (called ``kinks").
Both emit gravitational radiation in a highly concentrated beam.
Cusps are transient and produce a beam along a single direction, $\vu*e_\mathrm{c}$, while kinks propagate around the loop, beaming over a fanlike range of directions.
The cusp and kink waveforms are well approximated by~\cite{Damour:2001bk}
    \begin{align}
    \begin{split}
        \kappa_\mathrm{c}\qty(\nu_\mathrm{s},\vu*e_\mathrm{s})&\approx\frac{8}{\Gamma^2\qty(\frac{1}{3})}\qty(\frac{2}{3})^{2/3}\frac{G\mu l^{2/3}}{\nu_\mathrm{s}^{1/3}}\Theta\qty(\nu_\mathrm{s}-\frac{2}{l})\\
        &\qquad\qquad\qquad\qquad\times\Theta\qty(\theta_\mathrm{b}-\cos^{-1}\qty(\vu*e_\mathrm{s}\vdot\vu*e_\mathrm{c})),\\
        \kappa_\mathrm{k}\qty(\nu_\mathrm{s},\vu*e_\mathrm{s})&\approx\frac{2\sqrt{2}}{\uppi\Gamma\qty(\frac{1}{3})}\qty(\frac{2}{3})^{1/3}\frac{G\mu l^{1/3}}{\nu_\mathrm{s}^{2/3}}\Theta\qty(\nu_\mathrm{s}-\frac{2}{l})\\
        &\qquad\qquad\qquad\qquad\times\Theta\qty(\theta_\mathrm{b}-\cos^{-1}\qty(\vu*e_\mathrm{s}\vdot\vu*e_\mathrm{k})),
    \end{split}
    \end{align}
    where $\Gamma\qty(z)$ is the Euler gamma function, $\Theta\qty(x)$ is the Heaviside step function, $\vu*e_\mathrm{c}$ is the beaming direction of the cusp, and $\vu*e_\mathrm{k}$ is the direction closest to $\vu*e_\mathrm{s}$ within the ``fan".
Note that the the gravitational interaction of the strings is characterized by the dimensionless parameter $G\mu$, where $G$ is Newton's constant and $\mu$ the string tension.
The first step function reflects the fact that the GW frequency cannot be lower than the fundamental frequency of the loop, $2/l$.
The second step function ensures that the GW amplitude is zero outside the beam, with the beam opening angle given by
    \begin{equation}
        \theta_\mathrm{b}\approx\qty(\frac{4}{\sqrt{3}\nu_\mathrm{s}l})^{1/3}.
    \end{equation}
These different dependencies on $\vu*e_\mathrm{s}$ affect the integration over $\dd[2]{\sigma_\mathrm{s}}$.
For the cusp case, we choose spherical polar co\"ordinates $\qty(\theta_\mathrm{s},\phi_\mathrm{s})$ such that $\cos^{-1}\qty(\vu*e_\mathrm{s}\vdot\vu*e_\mathrm{c})=\theta_\mathrm{s}$.
Expanding in powers of $\theta_\mathrm{b}$, we find
    \begin{align*}
        \int_{S^2}\dd[2]{\sigma_\mathrm{s}}\Theta\qty(\theta_\mathrm{b}-\theta_\mathrm{s})&=2\uppi\int_0^{\theta_\mathrm{b}}\dd{\theta_\mathrm{s}}\sin\theta_\mathrm{s}\\
        &=\uppi\theta_\mathrm{b}^2+\order{\theta_\mathrm{b}^4}.
    \end{align*}
For the kink case, we approximate the fan as a great circle on the unit sphere.
This lets us choose co\"ordinates such that $\cos^{-1}\qty(\vu*e_\mathrm{s}\vdot\vu*e_\mathrm{k})=\qty|\theta_\mathrm{s}-\uppi/2|$, which gives
    \begin{align*}
        \int_{S^2}\dd[2]{\sigma_\mathrm{s}}\Theta\qty(\theta_\mathrm{b}-\qty|\theta_\mathrm{s}-\frac{\uppi}{2}|)&=2\uppi\int_{\frac{\uppi}{2}-\theta_\mathrm{b}}^{\frac{\uppi}{2}+\theta_\mathrm{b}}\dd{\theta_\mathrm{s}}\sin\theta_\mathrm{s}\\
        &=4\uppi\theta_\mathrm{b}+\order{\theta_\mathrm{b}^3}.
    \end{align*}
In both cases the observable signal is dominated by high frequencies $\nu_\mathrm{s}\gg1/l$.
This gives $\theta_\mathrm{b}^3\ll1$, so we neglect subleading terms in the above expressions.

In addition to cusps and kinks, collisions between propagating kinks might also be an important source of GW bursts~\cite{Binetruy:2010cc,Ringeval:2017eww}.
The radiation from these collisions is isotropic rather than beamed, and has a waveform
    \begin{equation}
        \kappa_\mathrm{kk}\qty(\nu_\mathrm{s})\approx\frac{G\mu}{\uppi^2\nu_\mathrm{s}}\Theta\qty(\nu_\mathrm{s}-\frac{2}{l}).
    \end{equation}
Kinks are created in pairs propagating in opposite directions along the loop, so the number of kink collisions per loop oscillation is
    \begin{equation}
        N_\mathrm{kk}=\frac{N_\mathrm{k}^2}{4}.
    \end{equation}
We therefore have
    \begin{align}
    \begin{split}
        \int_{S^2}\dd[2]{\sigma_\mathrm{s}}\kappa_\mathrm{c}^2&\approx A^2\qty(\nu_\mathrm{s}l)^{2/3}\frac{\qty(G\mu)^2}{\uppi^3\nu_\mathrm{s}^2}\Theta\qty(\nu_\mathrm{s}-\frac{2}{l}),\\
        \int_{S^2}\dd[2]{\sigma_\mathrm{s}}\kappa_\mathrm{k}^2&\approx 4A\qty(\nu_\mathrm{s}l)^{1/3}\frac{\qty(G\mu)^2}{\uppi^3\nu_\mathrm{s}^2}\Theta\qty(\nu_\mathrm{s}-\frac{2}{l}),\\
        \int_{S^2}\dd[2]{\sigma_\mathrm{s}}\kappa_\mathrm{kk}^2&\approx4\frac{\qty(G\mu)^2}{\uppi^3\nu_\mathrm{s}^2}\Theta\qty(\nu_\mathrm{s}-\frac{2}{l}),
    \end{split}
    \end{align}
    with $A$ a numerical constant, defined as
    \begin{equation}
        A\equiv\frac{2^{13/3}\uppi^2}{3^{5/6}\Gamma^2\qty(\frac{1}{3})}\approx11.0978
    \end{equation}
Using the above we can deduce the observable fraction of bursts of each type, $f_{\mathrm{o},i}$.
Let us write
    \begin{equation}
        f_{\mathrm{o},i}=f_{\mathrm{b},i}\Theta\qty(\nu_\mathrm{s}-\frac{2}{l}),
    \end{equation}
    where $f_{\mathrm{b},i}$ is the fraction of bursts that are beamed along the observer's past lightcone,
    \begin{align}
    \begin{split}
        f_\mathrm{b,c}&\approx\frac{\theta_\mathrm{b}^2}{4}\approx\qty(2\sqrt{3}\nu_\mathrm{s}l)^{-2/3},\\
        f_\mathrm{b,k}&\approx\theta_\mathrm{b}\approx\qty(\frac{\sqrt{3}\nu_\mathrm{s}l}{4})^{-1/3},\\
        f_\mathrm{b,kk}&=1.
    \end{split}
    \end{align}

\subsection{SGWB decomposition}
\label{eq:cs-sgwb-decomposition}
Summing the contributions from cusps, kinks, and kink-kink collisions and using Eq.~\eqref{eq:doppler-shift} to convert between $\nu_\mathrm{s}$ and $\nu_\mathrm{o}$, we obtain
    \begin{align}
    \begin{split}
        \Omega_\mathrm{gw}=&\frac{2\qty(G\mu)^2}{3\uppi^2H_\mathrm{o}^2\nu_\mathrm{o}}\int_0^{t_*}\frac{\dd{t}}{t^4}a^5\int_0^{\gamma_*}\frac{\dd{\gamma}}{\gamma}\bar{\mathcal{F}}\qty(1+\delta_\mathcal{F}+5\vu*e_\mathrm{o}\vdot\vb*v_\mathrm{o})\\
        &\times\Theta\qty(\gamma-\frac{2a}{\nu_\mathrm{o}t}\qty(1+\vu*e_\mathrm{o}\vdot\vb*v_\mathrm{o}))\\
        &\times\bigg[N_\mathrm{k}^2+4AN_\mathrm{k}\qty(1-\frac{1}{3}\vu*e_\mathrm{o}\vdot\vb*v_\mathrm{o})\qty(\frac{\nu_\mathrm{o}\gamma t}{a})^{1/3}\\
        &+A^2N_\mathrm{c}\qty(1-\frac{2}{3}\vu*e_\mathrm{o}\vdot\vb*v_\mathrm{o})\qty(\frac{\nu_\mathrm{o}\gamma t}{a})^{2/3}\bigg].
    \end{split}
    \end{align}
With reference to Sec.~\ref{sec:formalism-sgwb-decomposition}, we write this as
    \begin{align}
    \begin{split}
        \label{eq:cs-sgwb-x}
        \Omega_\mathrm{gw}=&\frac{2\qty(G\mu)^2}{3\uppi^2H_\mathrm{o}^2\nu_\mathrm{o}}\int_0^{t_*}\frac{\dd{t}}{t^4}a^5\int_0^{\gamma_*}\frac{\dd{\gamma}}{\gamma}\bar{\mathcal{F}}\qty(1+\delta_\mathcal{F})x^5\\
        &\times\Theta\qty(\gamma-\frac{2ax}{\nu_\mathrm{o}t})\\
        &\times\bigg[N_\mathrm{k}^2+4AN_\mathrm{k}\qty(\frac{\nu_\mathrm{o}\gamma t}{ax})^{1/3}\\
        &+A^2N_\mathrm{c}\qty(\frac{\nu_\mathrm{o}\gamma t}{ax})^{2/3}\bigg],
    \end{split}
    \end{align}
    where $x\equiv1+\vu*e_\mathrm{o}\vdot\vb*v_\mathrm{o}$ as before.
We therefore see that the averaged isotropic background value (monopole) is
    \begin{align}
    \begin{split}
        \label{eq:cs-monopole}
        \bar{\Omega}_\mathrm{gw}\equiv&\left.\Omega_\mathrm{gw}\right|_{x=1,\delta_\mathcal{F}=0}\\
        =&\frac{2\qty(G\mu)^2}{3\uppi^2H_\mathrm{o}^2\nu_\mathrm{o}}\int_0^{t_*}\frac{\dd{t}}{t^4}a^5\int_0^{\gamma_*}\frac{\dd{\gamma}}{\gamma}\bar{\mathcal{F}}\Theta\qty(\gamma-\frac{2a}{\nu_\mathrm{o}t})\\
        &\times\qty[N_\mathrm{k}^2+4AN_\mathrm{k}\qty(\frac{\nu_\mathrm{o}\gamma t}{a})^{1/3}+A^2N_\mathrm{c}\qty(\frac{\nu_\mathrm{o}\gamma t}{a})^{2/3}],
    \end{split}
    \end{align}
    with the source anisotropies given by
    \begin{align}
    \begin{split}
        \delta_\mathrm{gw}^\qty(\mathrm{s})\equiv&\left.\delta_\mathrm{gw}\right|_{x=1}=\frac{\left.\Omega_\mathrm{gw}\right|_{x=1}-\bar{\Omega}_\mathrm{gw}}{\bar{\Omega}_\mathrm{gw}}\\
        =&\bar{\Omega}_\mathrm{gw}^{-1}\frac{2\qty(G\mu)^2}{3\uppi^2H_\mathrm{o}^2\nu_\mathrm{o}}\int_0^{t_*}\frac{\dd{t}}{t^4}a^5\int_0^{\gamma_*}\frac{\dd{\gamma}}{\gamma}\bar{\mathcal{F}}\delta_\mathcal{F}\Theta\qty(\gamma-\frac{2a}{\nu_\mathrm{o}t})\\
        &\times\qty[N_\mathrm{k}^2+4AN_\mathrm{k}\qty(\frac{\nu_\mathrm{o}\gamma t}{a})^{1/3}+A^2N_\mathrm{c}\qty(\frac{\nu_\mathrm{o}\gamma t}{a})^{2/3}].
    \end{split}
    \end{align}
The dipole factor is straightforward to evaluate from Eqs.~\eqref{eq:dipole-factor} and \eqref{eq:cs-sgwb-x}, noting that $\pdv{x}\Theta\qty(\gamma-\frac{2ax}{\nu_\mathrm{o}t})=-\frac{2a}{\nu_\mathrm{o}t}\delta\qty(\gamma-\frac{2ax}{\nu_\mathrm{o}t})$.
We therefore have
    \begin{align}
    \begin{split}
        \mathcal{D}=&v_\mathrm{o}\bar{\Omega}_\mathrm{gw}^{-1}\left.\pdv{\Omega_\mathrm{gw}}{x}\right|_{x=1,\delta_\mathcal{F}=0}\\
        =&v_\mathrm{o}\bar{\Omega}_\mathrm{gw}^{-1}\frac{2\qty(G\mu)^2}{9\uppi^2H_\mathrm{o}^2\nu_\mathrm{o}}\int_0^{t_*}\frac{\dd{t}}{t^4}a^5\Bigg\{\int_0^{\gamma_*}\frac{\dd{\gamma}}{\gamma}\bar{\mathcal{F}}\qty(\gamma)\Theta\qty(\gamma-\frac{2a}{\nu_\mathrm{o}t})\\
        &\times\qty[15N_\mathrm{k}^2+56AN_\mathrm{k}\qty(\frac{\nu_\mathrm{o}\gamma t}{a})^{1/3}+13N_\mathrm{c}^2\qty(\frac{\nu_\mathrm{o}\gamma t}{a})^{2/3}]\\
        &-3\qty(N_\mathrm{k}^2+2^{7/3}AN_\mathrm{k}+2^{2/3}A^2N_\mathrm{c})\bar{\mathcal{F}}\qty(\frac{2a}{\nu_\mathrm{o}t})\Bigg\}.
    \end{split}
    \end{align}

\subsection{Matter and radiation eras}
In order to evaluate the integrals in the expressions above, we consider the contributions from the matter era (ME) and radiation era (RE) separately.
We define the dimensionless parameters
	\begin{equation}
		\tau\equiv\frac{t}{t_\mathrm{o}},\qquad\omega\equiv t_\mathrm{o}\nu_\mathrm{o},
	\end{equation}
    so that $\tau\in\qty[0,1]$ and $\omega\gg1$ (since any GW frequency we can observe is much larger than the Hubble frequency).
The scale factor can be approximated by
	\begin{equation}
		a(\tau)=\left\{
        \begin{matrix*}[l]
            a_\mathrm{eq}^{1/4}\tau^{1/2},&0\le\tau<a_\mathrm{eq}^{3/2}\quad\text{(RE)}\\
            \tau^{2/3},&a_\mathrm{eq}^{3/2}\le\tau\le1\quad\text{(ME)}
        \end{matrix*}
        \right.
	\end{equation}
    where $a_\mathrm{eq}$ is the scale factor at matter-radiation equality.
This gives
    \begin{equation}
        \gamma_*\qty(\tau)=\left\{
        \begin{matrix*}[l]
            2,&0\le\tau<a_\mathrm{eq}^{3/2}\quad\text{(RE)}\\
            3-a_\mathrm{eq}^{1/2}\tau^{-1/3},&a_\mathrm{eq}^{3/2}\le\tau\le1\quad\text{(ME)}
        \end{matrix*}
        \right.
    \end{equation}
    \begin{equation}
        r\qty(\tau)=\left\{
        \begin{matrix*}[l]
            3t_\mathrm{o}\qty(1-\frac{a_\mathrm{eq}^{3/4}+2\tau^{1/2}}{3a_\mathrm{eq}^{1/4}}),&0\le\tau<a_\mathrm{eq}^{3/2}\quad\text{(RE)}\\
            3t_\mathrm{o}\qty(1-\tau^{1/3}),&a_\mathrm{eq}^{3/2}\le\tau\le1\quad\text{(ME)}
        \end{matrix*}
        \right.
    \end{equation}
Although the background distribution $\bar{\mathcal{F}}$ is constant in time during each era, it usually differs between eras, so we let
    \begin{equation}
        \bar{\mathcal{F}}\qty(\gamma)=\left\{
        \begin{matrix*}[l]
            \bar{\mathcal{F}}_\mathrm{r}\qty(\gamma),&0\le\tau<a_\mathrm{eq}^{3/2}\quad\text{(RE)}\\
            \bar{\mathcal{F}}_\mathrm{m}\qty(\gamma),&a_\mathrm{eq}^{3/2}\le\tau\le1\quad\text{(ME)}
        \end{matrix*}
        \right.
    \end{equation}
We can manipulate the step function by altering the lower limits of the integrals, e.g. $\int_0^{\gamma_*}\dd{\gamma}\Theta\qty(\gamma-\frac{2a}{\omega\tau})=\Theta\qty(\gamma_*-\frac{2a}{\omega\tau})\int_{2a/\omega\tau}^{\gamma_*}\dd{\gamma}$.
Recalling that $\omega\gg1$, we see that the step function $\Theta\qty(\gamma_*-\frac{2a}{\omega\tau})$ is only zero when $\tau\ll1$, i.e.~at the beginning of the radiation era.
Working this through, we find that the monopole and kinematic dipole are given by
    \begin{equation}
        \bar{\Omega}_\mathrm{gw}=\frac{\qty(G\mu)^2}{\omega}\qty(N_\mathrm{k}^2I_{\bar{\Omega}}^\qty(0)+4AN_\mathrm{k}I_{\bar{\Omega}}^\qty(1/3)\omega^{1/3}+A^2N_\mathrm{c}I_{\bar{\Omega}}^\qty(2/3)\omega^{2/3}),
    \end{equation}
    \begin{align}
    \begin{split}
        \mathcal{D}=&v_\mathrm{o}\bar{\Omega}_\mathrm{gw}^{-1}\frac{\qty(G\mu)^2}{\omega}\bigg[5N_\mathrm{k}^2I_{\bar{\Omega}}^\qty(0)+\frac{56}{3}AN_\mathrm{k}I_{\bar{\Omega}}^\qty(1/3)\omega^{1/3}\\
        &+\frac{13}{3}A^2N_\mathrm{c}I_{\bar{\Omega}}^\qty(2/3)\omega^{2/3}-(N_\mathrm{k}^2+2^{7/3}AN_\mathrm{k}\\
        &+2^{2/3}A^2N_\mathrm{c})I_\mathcal{D}\bigg],
    \end{split}
    \end{align}
 respectively, where we define the integrals
    \begin{align}
    \begin{split}
        \label{eq:monopole-integrals}
        I_{\bar{\Omega}}^\qty(q)\equiv&\frac{2a_\mathrm{eq}^{\frac{5-q}{4}}}{3\qty(\uppi H_\mathrm{o}t_\mathrm{o})^2}\int^{\tau_{**}}_\frac{a_\mathrm{eq}^{1/2}}{\omega^2}\frac{\dd{\tau}}{\tau^\frac{3-q}{2}}\int^2_\frac{2a_\mathrm{eq}^{1/4}}{\omega\tau^{1/2}}\frac{\dd{\gamma}}{\gamma^{1-q}}\bar{\mathcal{F}}_\mathrm{r}\Theta\qty(\tau_{**}-\frac{a_\mathrm{eq}^{1/2}}{\omega^2})\\
        &+\frac{2}{3\qty(\uppi H_\mathrm{o}t_\mathrm{o})^2}\int^{\tau_*}_{a_\mathrm{eq}^{3/2}}\frac{\dd{\tau}}{\tau^\frac{2-q}{3}}\int^{3-\frac{a_\mathrm{eq}^{1/2}}{\tau^{1/3}}}_{2/\omega\tau^{1/3}}\frac{\dd{\gamma}}{\gamma^{1-q}}\bar{\mathcal{F}}_\mathrm{m}\Theta\qty(\tau_*-a_\mathrm{eq}^{3/2}),
    \end{split}\\
    \begin{split}
        \label{eq:dipole-integrals}
        I_\mathcal{D}\equiv&\frac{2a_\mathrm{eq}^{5/4}}{3\qty(\uppi H_\mathrm{o}t_\mathrm{o})^2}\int^{\tau_{**}}_0\frac{\dd{\tau}}{\tau^{3/2}}\bar{\mathcal{F}}_\mathrm{r}\qty(\frac{2a_\mathrm{eq}^{1/4}}{\omega\tau^{1/2}})\Theta\qty(\tau_{**}-\frac{a_\mathrm{eq}^{1/2}}{\omega^2})\\
        &+\frac{2}{3\qty(\uppi H_\mathrm{o}t_\mathrm{o})^2}\int^{\tau_*}_{a_\mathrm{eq}^{3/2}}\frac{\dd{\tau}}{\tau^{2/3}}\bar{\mathcal{F}}_\mathrm{m}\qty(\frac{2}{\omega\tau^{1/3}})\Theta\qty(\tau_*-a_\mathrm{eq}^{3/2}),
    \end{split}
    \end{align}
with the upper limit for the radiation era being given by
    \begin{equation}
        \tau_{**}\equiv\min\qty(a_\mathrm{eq}^{3/2},\tau_*).
    \end{equation}

In order to calculate $\tau_*$, we solve the integral equation $\Lambda\qty(\tau_*,\omega)=1$ for each frequency $\omega$.
The duty cycle is now given by
    \begin{equation}
        \label{eq:duty-cycle}
        \Lambda\qty(\tau,\omega)=\frac{N_\mathrm{k}^2I_{\Lambda}^\qty(0)}{\omega}+\frac{4N_\mathrm{k}I_{\Lambda}^\qty(1/3)}{\omega^{4/3}}+\frac{N_\mathrm{c}I_{\Lambda}^\qty(2/3)}{\omega^{5/3}},
    \end{equation}
    where we define another family of integrals,
    \begin{align}
    \begin{split}
        \label{eq:lambda-integrals}
        I_{\Lambda}^\qty(q)\qty(\tau,\omega)\equiv&2^{1+2q}3^\frac{4-q}{2}\uppi a_\mathrm{eq}^{1/2}\int_{\max\qty(\tau,\frac{a_\mathrm{eq}^{1/2}}{\omega^2})}^{a_\mathrm{eq}^{3/2}}\frac{\dd{\tau'}}{\tau'^{\frac{9+q}{3}}}\\
        &\times\qty(1-\frac{a_\mathrm{eq}^{3/4}+2\tau'^{1/2}}{3a_\mathrm{eq}^{1/4}})^2\int_{\frac{2a_\mathrm{eq}^{1/4}}{\omega\tau'^{1/2}}}^2\frac{\dd{\gamma}}{\gamma^{1+q}}\mathcal{F}_\mathrm{r}\Theta\qty(a_\mathrm{eq}^{3/2}-\tau)\\
        &+2^{1+2q}3^\frac{4-q}{2}\uppi\int_{\max(\tau,a_\mathrm{eq}^{3/2})}^1\frac{\dd{\tau'}}{\tau'^{\frac{8+q}{3}}}\qty(1-\tau'^{1/3})^2\\
        &\times\int_{\frac{2}{\omega\tau'^{1/3}}}^{3-\frac{a_\mathrm{eq}^{1/2}}{\tau'^{1/3}}}\frac{\dd{\gamma}}{\gamma^{1+q}}\mathcal{F}_\mathrm{m}.
    \end{split}
    \end{align}
If $\Lambda(\tau, \omega)<1$ for all $\tau$, then we define $\tau_*(\omega)=0$.

In principle, all we now need to calculate $\bar{\Omega}_\mathrm{gw}$ and $\mathcal{D}$ is the homogeneous loop distribution function $\bar{\mathcal{F}}\qty(\gamma)$.
For $\delta_\mathrm{gw}^\qty(\mathrm{s})$ however, we need to know the density contrast $\delta_\mathcal{F}$; this is addressed in the following section.

\subsection{Two-point correlation function}
\label{sec:2pcf}
We are not able to map out $\delta_\mathcal{F}$ by observing string loops directly, so instead we treat it statistically.
Following Ref.~\cite{Olmez:2011cg}, we consider anisotropies in the SGWB produced by random fluctuations in the number of GW sources, leading to correlations between different directions in the sky, expressed in terms of the two-point correlation function (2PCF)
    \begin{equation}
        C_\mathrm{gw}\qty(\theta_\mathrm{o},\nu_\mathrm{o})\equiv\ev{\delta_\mathrm{gw}^\qty(\mathrm{s})\qty(\nu_\mathrm{o},\vu*e_\mathrm{o})\delta_\mathrm{gw}^\qty(\mathrm{s})(\nu_\mathrm{o},\vu*e\mathrlap{'}_\mathrm{o})},
    \end{equation}
    where as before $\theta_\mathrm{o}\equiv\cos^{-1}(\vu*e_\mathrm{o}\vdot\vu*e\mathrlap{'}_\mathrm{o})$, and the angle brackets denote an averaging over all pairs of directions $\vu*e_\mathrm{o}$, $\vu*e\mathrlap{'}_\mathrm{o}$ whose angle of separation is $\theta_\mathrm{o}$.

We can write the SGWB monopole as
    \begin{equation}
        \bar{\Omega}_\mathrm{gw}\qty(\nu_\mathrm{o})=\int_0^{t_*}\frac{\dd[3]{V\qty(t)}}{\dd[2]{\sigma_\mathrm{o}}}\int_0^{\gamma_*t}\dd{l}\bar{n}w,
    \end{equation}
    where $\dd[3]{V}=\dd[2]{\sigma_\mathrm{o}}\dd{t}a^2r^2$.
Here $w\qty(\nu_\mathrm{o},t,l)$ is the average energy contribution per loop and $\bar{n}\qty(t,l)$ is the isotropic loop number density, defined such that $\dd{t}\dd{l}\bar{n}a^2r^2$ is the average number of sources per unit solid angle at times between $t$ and $t+\dd{t}$ on the observer's past light cone, with length between $l$ and $l+\dd{l}$---this ensures that the function $\bar{n}$ is the same as that used previously.

Now we let the number of loops have random Poisson-like fluctuations (as one would expect for any large number of discrete objects), and assume that these fluctuations are only correlated over small angular scales, and only at equal times $t$ for loops with equal sizes $l$.
Then, using the results found in Ref.~\cite{Olmez:2011cg}, we find that the 2PCF of $\delta_\mathrm{gw}^\qty(\mathrm{s})$ is given by
    \begin{equation}
        \label{eq:olmez-mandic-siemens-2pcf}
        C_\mathrm{gw}\approx\bar{\Omega}_\mathrm{gw}^{-2}\int_0^{t_*}\dd{t}a^2r^2\int_0^{\gamma_*t}\dd{l}\bar{n}w^2\mathcal{C},
    \end{equation}
    where the function $\mathcal{C}\qty(\theta_\mathrm{o},t,l)$ encodes the angular correlation of loops with length $l$ at time $t$.
Rewriting Eq.~\eqref{eq:cs-monopole} in terms of $l$ and $t$, it reads
    \begin{align}
    \begin{split}
        \bar{\Omega}_\mathrm{gw}=&\frac{2\qty(G\mu)^2}{3\uppi^2H_\mathrm{o}^2\nu_\mathrm{o}}\int_0^{t_*}\dd{t}a^5\int_0^{\gamma_*t}\frac{\dd{l}}{l}\bar{n}\Theta\qty(\nu_\mathrm{o}l-2a)\\
        &\times\qty[N_\mathrm{k}^2+4AN_\mathrm{k}\qty(\frac{\nu_\mathrm{o}l}{a})^{1/3}+A^2N_\mathrm{c}\qty(\frac{\nu_\mathrm{o}l}{a})^{2/3}],
    \end{split}
    \end{align}
    so by comparison, we deduce that
    \begin{align}
    \begin{split}
        w=&\frac{2\qty(G\mu)^2}{3\uppi^2H_\mathrm{o}^2\nu_\mathrm{o}}\frac{a^3}{r^2l}\Theta\qty(\nu_\mathrm{o}l-2a)\bigg[N_\mathrm{k}^2+4AN_\mathrm{k}\qty(\frac{\nu_\mathrm{o}l}{a})^{1/3}\\
        &+A^2N_\mathrm{c}\qty(\frac{\nu_\mathrm{o}l}{a})^{2/3}\bigg].
    \end{split}
    \end{align}
Using Eq.~\eqref{eq:olmez-mandic-siemens-2pcf}, the 2PCF is therefore given by
    \begin{align}
    \begin{split}
        C_\mathrm{gw}=&\bar{\Omega}_\mathrm{gw}^{-2}\frac{4\qty(G\mu)^4}{9\uppi^4H_\mathrm{o}^4\nu_\mathrm{o}^2}\int_0^{t_*}\dd{t}\frac{a^8}{t^5r^2}\int_0^{\gamma_*}\frac{\dd{\gamma}}{\gamma^2}\bar{\mathcal{F}}\mathcal{C}\Theta\qty(\gamma-\frac{2a}{\nu_\mathrm{o}t})\\
        &\times\bigg[N_\mathrm{k}^4+8AN_\mathrm{k}^3\qty(\frac{\nu_\mathrm{o}t\gamma}{a})^\frac{1}{3}+2A^2N_\mathrm{k}^2\qty(N_\mathrm{c}+8)\qty(\frac{\nu_\mathrm{o}t\gamma}{a})^\frac{2}{3}\\
        &+8A^3N_\mathrm{k}N_\mathrm{c}\qty(\frac{\nu_\mathrm{o}t\gamma}{a})+A^4N_\mathrm{c}^2\qty(\frac{\nu_\mathrm{o}t\gamma}{a})^\frac{4}{3}\bigg].
    \end{split}
    \end{align}
All that remains is to determine $\mathcal{C}$.
Suppose that there is a characteristic length scale over which loops are correlated.
We expect this correlation length to scale with the loop size $l$, so we take it to be $kl$, where $k$ is an ``ignorance factor" of order unity.
This translates into a sky angle
    \begin{equation}
        \theta_\mathcal{C}\qty(t,\gamma)=2\tan^{-1}\qty(\frac{k\gamma t}{ar}),
    \end{equation}
which is the maximum angular size of any correlated features.
On smaller scales than $ \theta_\mathcal{C}$, the 2PCF measures the relative local size of the number density contrast (which is set by the size of the Poisson fluctuations).
On larger scales than $\theta_\mathcal{C}$, the 2PCF measures the global size of the density contrast, which is zero by definition.
We therefore write
    \begin{equation}
        \mathcal{C}\qty(\theta_\mathrm{o},t,\gamma)\equiv\Theta\qty(\theta_\mathcal{C}-\theta_\mathrm{o})=\Theta\qty(\frac{k\gamma t}{ar}-\tan\frac{\theta_\mathrm{o}}{2}),
    \end{equation}
    which implies
    \begin{align}
    \begin{split}
        \label{eq:final-Cgw}
        C_\mathrm{gw}=&\bar{\Omega}_\mathrm{gw}^{-2}\frac{4\qty(G\mu)^4}{9\uppi^4H_\mathrm{o}^4\nu_\mathrm{o}^2}\int_0^{t_*}\dd{t}\frac{a^8}{t^5r^2}\int_0^{\gamma_*}\frac{\dd{\gamma}}{\gamma^2}\bar{\mathcal{F}}\Theta\qty(\gamma-\frac{2a}{\nu_\mathrm{o}t})\\
        &\times\Theta\qty(\gamma-\frac{ar}{kt}\tan\frac{\theta_\mathrm{o}}{2})\bigg[N_\mathrm{k}^4+8AN_\mathrm{k}^3\qty(\frac{\nu_\mathrm{o}t\gamma}{a})^\frac{1}{3}\\
        &+2A^2N_\mathrm{k}^2\qty(N_\mathrm{c}+8)\qty(\frac{\nu_\mathrm{o}t\gamma}{a})^\frac{2}{3}+8A^3N_\mathrm{k}N_\mathrm{c}\qty(\frac{\nu_\mathrm{o}t\gamma}{a})\\
        &+A^4N_\mathrm{c}^2\qty(\frac{\nu_\mathrm{o}t\gamma}{a})^\frac{4}{3}\bigg].
    \end{split}
    \end{align}
Equation~\eqref{eq:final-Cgw} is the second main result of our study.
For any model of cosmic strings for which the loop distribution is known, one can use Eq.~\eqref{eq:final-Cgw} to calculate the correlation function of the resulting SGWB, and therefore fully describe its anisotropies.

Evaluating Eq.~\eqref{eq:final-Cgw} analytically for all $\nu_\mathrm{o}$ and $\theta_\mathrm{o}$ is made considerably more difficult by the two competing step functions.
However, we are only interested in loops whose proper distance from us is greater than $r_\mathrm{min}\equiv r\qty(t_*)$.
We can limit ourselves to the region of the $\nu_\mathrm{o}$-$\theta_\mathrm{o}$ parameter space in which $\Theta\qty(\gamma-\frac{ar}{kt}\tan\frac{\theta_\mathrm{o}}{2})$ is always stricter than the step function $\Theta\qty(\gamma-\frac{2a}{\nu_\mathrm{o}t})$, leading to the constraint
    \begin{equation}
        \theta_\mathrm{o}\ge\theta_\mathrm{min}\equiv2\tan^{-1}\qty(\frac{2k}{\nu_\mathrm{o}r_\mathrm{min}}).
    \end{equation}
We expect $r_\mathrm{min}$ to be no smaller than a few orders of magnitude less than the Hubble length, and $\nu_\mathrm{o}$ to be many orders of magnitude greater than the Hubble frequency, so $\nu_\mathrm{o}r_\mathrm{min}\gg1$, and $\theta_\mathrm{min}\ll1$.
\begin{figure*}[t]
    \includegraphics[width=\textwidth]{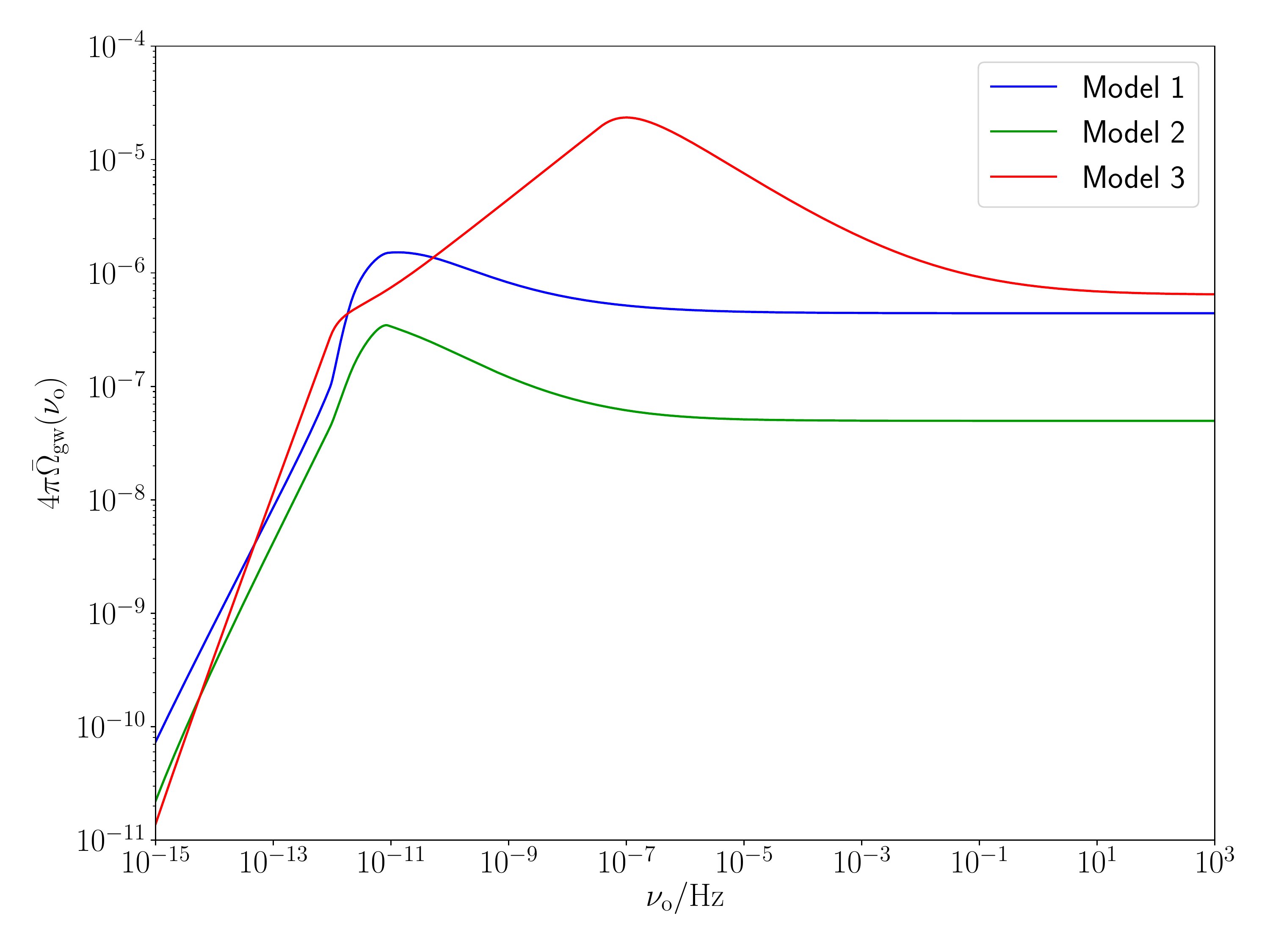}
    \caption{The frequency spectrum of the SGWB monopole $\bar{\Omega}_\mathrm{gw}\qty(\nu_\mathrm{o})$ in each of the three models for the loop distribution, using $G\mu=10^{-7}$, and $N_\mathrm{c}=N_\mathrm{k}=1$.}
    \label{fig:monopole-models}
\end{figure*}

Evaluating the integrals, we find that the correlation of points separated by angles $\theta_\mathrm{o}\ge\theta_\mathrm{min}$ is
    \begin{align}
    \begin{split}
        \label{eq:final-Cgw-integrals}
        C_\mathrm{gw}=&\bar{\Omega}_\mathrm{gw}^{-2}\frac{\qty(G\mu)^4}{\omega^2}\bigg[N_\mathrm{k}^4I_C^\qty(0)+8AN_\mathrm{k}^3I_C^\qty(1/3)\omega^{1/3}\\
        &+2A^2N_\mathrm{k}^2\qty(N_\mathrm{c}+8)I_C^\qty(2/3)\omega^{2/3}\\
        &+8A^3N_\mathrm{k}N_\mathrm{c}I_C^\qty(1)\omega+A^4N_\mathrm{c}^2I_C^\qty(4/3)\omega^{4/3}\bigg],
    \end{split}
    \end{align}
    where we define the integrals
    \begin{align}
    \begin{split}
        I_C^\qty(q)\qty(\theta_\mathrm{o},\omega)\equiv&\frac{4a_\mathrm{eq}^\frac{8-q}{4}}{9\qty(\uppi H_\mathrm{o}t_\mathrm{o})^4}\int_{\tau_\mathrm{r}}^{\tau_{**}}\frac{\dd{\tau}}{\tau^\frac{2-q}{2}}\qty(1-\frac{a_\mathrm{eq}^{3/4}+2\tau^{1/2}}{3a_\mathrm{eq}^{1/4}})^{-2}\\
        &\times\int_{\gamma_\mathrm{r}}^2\frac{\dd{\gamma}}{\gamma^{2-q}}\bar{\mathcal{F}}_\mathrm{r}\Theta\qty(\theta_\mathrm{r}^*-\theta_\mathrm{o})+\frac{4}{9\qty(\uppi H_\mathrm{o}t_\mathrm{o})^4}\\
        &\times\int_{\tau_\mathrm{m}}^{\tau_*}\dd{\tau}\frac{\tau^\frac{1+q}{3}}{\qty(1-\tau^{1/3})^2}\int_{\gamma_\mathrm{m}}^{3-\frac{a_\mathrm{eq}^{1/2}}{\tau^{1/3}}}\frac{\dd{\gamma}}{\gamma^{2-q}}\\
        &\times\bar{\mathcal{F}}_\mathrm{m}\Theta\qty(\theta_\mathrm{m}^*-\theta_\mathrm{o}),
    \end{split}
    \end{align}
    with $\tau_{**}\equiv\min\qty(a_\mathrm{eq}^{3/2},\tau_*)$ as before, and with further limits defined by
\begin{widetext}
    \begin{align}
    \begin{split}
        \label{eq:theta-eq}
        \theta_\mathrm{eq}&\equiv2\tan^{-1}\qty[\frac{2a_\mathrm{eq}^{1/2}k}{3\qty(1-a_\mathrm{eq}^{1/2})}],\qquad\theta_\mathrm{m}^*\equiv2\tan^{-1}\qty[k\frac{\tau_*^{1/3}-\frac{1}{3}a_\mathrm{eq}^{1/2}}{1-\tau_*^{1/3}}],\\
        \theta_\mathrm{r}^*&\equiv2\tan^{-1}\qty[\frac{2k\tau_*^{1/2}}{a_\mathrm{eq}^{1/4}\qty(3-a_\mathrm{eq}^{1/2})-2\tau_*^{1/2}}]\Theta\qty(a_\mathrm{eq}^{3/2}-\tau_*)+\theta_\mathrm{eq}\Theta\qty(\tau_*-a_\mathrm{eq}^{3/2}),\\
        \tau_\mathrm{r}\qty(\theta_\text{o})&\equiv\frac{a_\mathrm{eq}^{1/2}}{4}\qty(3-a_\mathrm{eq}^{1/2})^2\qty(\frac{\tan\frac{\theta_\mathrm{o}}{2}}{k+\tan\frac{\theta_\mathrm{o}}{2}})^2,\qquad\tau_\mathrm{m}\qty(\theta_\text{o})\equiv\frac{1}{27}\qty(\frac{a_\mathrm{eq}^{1/2}k+3\tan\frac{\theta_\mathrm{o}}{2}}{k+\tan\frac{\theta_\mathrm{o}}{2}})^3\Theta\qty(\theta_\mathrm{o}-\theta_\mathrm{eq})+a_\mathrm{eq}^{3/2}\Theta\qty(\theta_\mathrm{eq}-\theta_\mathrm{o}),\\
        \gamma_\mathrm{r}\qty(\theta_\text{o},\tau)&\equiv\frac{3a_\mathrm{eq}^{1/4}}{k\tau^{1/2}}\qty(1-\frac{a_\mathrm{eq}^{3/4}+2\tau^{1/2}}{3a_\mathrm{eq}^{1/4}})\tan\frac{\theta_\mathrm{o}}{2},\qquad\gamma_\mathrm{m}\qty(\theta_\text{o},\tau)\equiv\frac{3}{k\tau^{1/3}}\qty(1-\tau^{1/3})\tan\frac{\theta_\mathrm{o}}{2}.
    \end{split}
    \end{align}
Let us emphasize that Eq.~\eqref{eq:final-Cgw-integrals} is valid only for angles $\theta_\mathrm{o}\ge\theta_\mathrm{min}$, where
    \begin{equation}
        \label{eq:theta-min}
        \theta_\mathrm{min}=2\tan^{-1}\!\qty[\frac{2k}{3\omega}\qty(1-\frac{a_\mathrm{eq}^{3/4}+2\tau_*^{1/2}}{3a_\mathrm{eq}^{1/4}})^{-1}]\Theta\qty(a_\mathrm{eq}^{3/2}-\tau_*)+2\tan^{-1}\qty[\frac{2k}{3\omega}\qty(1-\tau_*^{1/3})^{-1}]\Theta\qty(\tau_*-a_\mathrm{eq}^{3/2}).
    \end{equation}
We can rewrite this in a form that simplifies the limits by replacing $\theta_\mathrm{o}$ with $u\equiv\tan\frac{\theta_\mathrm{o}}{2}$.
This gives
    \begin{align}
    \begin{split}
        \label{eq:2pcf-u-integrals}
        I_C^\qty(q)\qty(\theta_\mathrm{o},\omega)\equiv\frac{4a_\mathrm{eq}^\frac{8-q}{4}}{9\qty(\uppi H_\mathrm{o}t_\mathrm{o})^4}&\int_{\tau_\mathrm{r}}^{\tau_{**}}\frac{\dd{\tau}}{\tau^\frac{2-q}{2}}\qty(1-\frac{a_\mathrm{eq}^{3/4}+2\tau^{1/2}}{3a_\mathrm{eq}^{1/4}})^{-2}\int_{\gamma_\mathrm{r}}^2\frac{\dd{\gamma}}{\gamma^{2-q}}\bar{\mathcal{F}}_\mathrm{r}\Theta\qty(u_\mathrm{r}^*-u)\\
        &+\frac{4}{9\qty(\uppi H_\mathrm{o}t_\mathrm{o})^4}\int_{\tau_\mathrm{m}}^{\tau_*}\dd{\tau}\frac{\tau^\frac{1+q}{3}}{\qty(1-\tau^{1/3})^2}\int_{\gamma_\mathrm{m}}^{3-\frac{a_\mathrm{eq}^{1/2}}{\tau^{1/3}}}\frac{\dd{\gamma}}{\gamma^{2-q}}\bar{\mathcal{F}}_\mathrm{m}\Theta\qty(u_\mathrm{m}^*-u),
    \end{split}
    \end{align}
    where the limits are rewritten as
    \begin{align}
    \begin{split}
        \label{eq:2pcf-u-limits}
        u_\mathrm{eq}&\equiv\frac{2a_\mathrm{eq}^{1/2}k}{3\qty(1-a_\mathrm{eq}^{1/2})},\qquad u_\mathrm{m}^*\equiv k\frac{\tau_*^{1/3}-\frac{1}{3}a_\mathrm{eq}^{1/2}}{1-\tau_*^{1/3}},\qquad u_\mathrm{r}^*\equiv\frac{2k\tau_*^{1/2}}{a_\mathrm{eq}^{1/4}\qty(3-a_\mathrm{eq}^{1/2})-2\tau_*^{1/2}}\Theta\qty(a_\mathrm{eq}^{3/2}-\tau_*)+u_\mathrm{eq}\Theta\qty(\tau_*-a_\mathrm{eq}^{3/2}),\\
        \tau_\mathrm{r}\qty(u)&\equiv\frac{a_\mathrm{eq}^{1/2}}{4}\qty(3-a_\mathrm{eq}^{1/2})^2\qty(\frac{u}{k+u})^2,\qquad\tau_\mathrm{m}\qty(u)\equiv\frac{1}{27}\qty(\frac{a_\mathrm{eq}^{1/2}k+3u}{k+u})^3\Theta\qty(u-u_\mathrm{eq})+a_\mathrm{eq}^{3/2}\Theta\qty(u_\mathrm{eq}-u),\\
        \gamma_\mathrm{r}\qty(u,\tau)&\equiv\frac{3a_\mathrm{eq}^{1/4}u}{k\tau^{1/2}}\qty(1-\frac{a_\mathrm{eq}^{3/4}+2\tau^{1/2}}{3a_\mathrm{eq}^{1/4}}),\qquad\gamma_\mathrm{m}\qty(u,\tau)\equiv\frac{3u}{k\tau^{1/3}}\qty(1-\tau^{1/3}).
    \end{split}
    \end{align}
\end{widetext}

\subsection{Calculating the \texorpdfstring{$C_\ell$}{} spectrum}
\label{sec:C_l}
\begin{figure*}[t]
    \includegraphics[width=\textwidth]{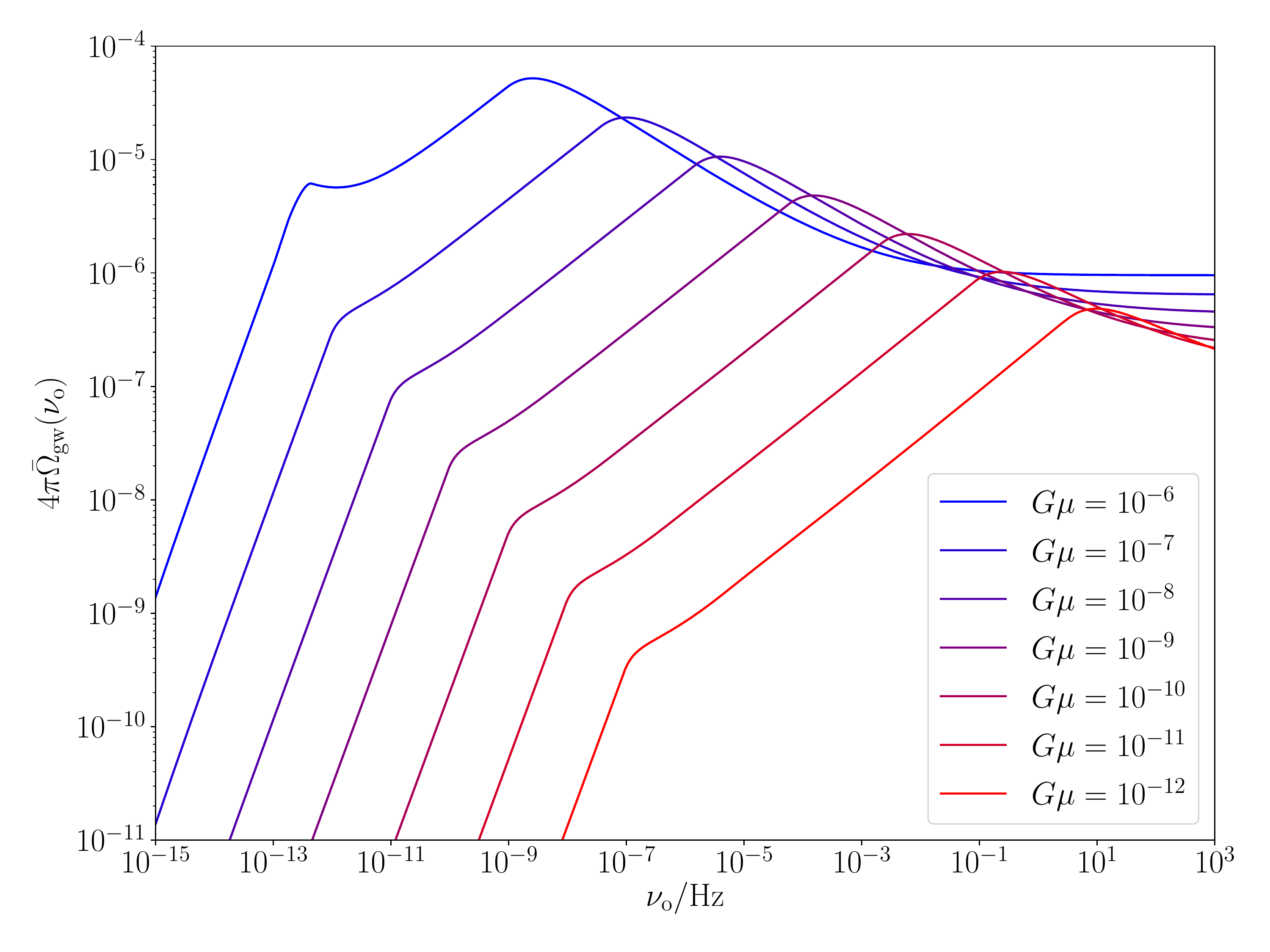}
    \caption{The frequency spectrum of the SGWB monopole $\bar{\Omega}_\mathrm{gw}\qty(\nu_\mathrm{o})$ in model 3 for a range of values of $G\mu$, with $N_\mathrm{c}=N_\mathrm{k}=1$.}
    \label{fig:monopole-Gmu}
\end{figure*}
The 2PCF is usually expanded in terms of Legendre polynomials using Eq.~\eqref{eq:multipole-expansion}, where the coefficients $C_\ell$ can be thought of as describing the statistics of $\Omega_\mathrm{gw}$ on angular scales $\uppi/\ell$.
The combination $\ell\qty(\ell+1)C_\ell/2\uppi$ is roughly the contribution to the anisotropic variance of $\delta^{(\mathrm{s})}_\mathrm{gw}$ per logarithmic bin in $\ell$.
Using the orthonormality condition for the Legendre polynomials,
    \begin{equation}
        \label{eq:legendre-orthogonal}
        \int_{-1}^{+1}\dd{x}P_\ell\qty(x)P_{\ell'}\qty(x)=\frac{2}{2\ell+1}\delta_{\ell\ell'},
    \end{equation}
    we write
    \begin{align*}
        C_\ell&=2\uppi\int_{-1}^{+1}\dd(\cos\theta_\mathrm{o})P_\ell\qty(\cos\theta_\mathrm{o})C_\mathrm{gw}\qty(\theta_\mathrm{o},\omega)\\
        &=2\uppi\int_0^\uppi\dd{\theta_\mathrm{o}}\sin\theta_\mathrm{o} P_\ell\qty(\cos\theta_\mathrm{o})C_\mathrm{gw}\qty(\theta_\mathrm{o},\omega).
    \end{align*}
Inserting our expression Eq.~\eqref{eq:final-Cgw-integrals} for $C_\mathrm{gw}$, we therefore find
    \begin{align}
    \begin{split}
        \label{eq:C_l-cs}
        C_\ell=\bar{\Omega}_\mathrm{gw}^{-2}\frac{\qty(G\mu)^4}{\omega^2}\bigg[&N_\mathrm{k}^4I_\ell^\qty(0)+8AN_\mathrm{k}^3I_\ell^\qty(1/3)\omega^{1/3}\\
        &+2A^2N_\mathrm{k}^2\qty(N_\mathrm{c}+8)I_\ell^\qty(2/3)\omega^{2/3}\\
        &+8A^3N_\mathrm{k}N_\mathrm{c}I_\ell^\qty(1)\omega+A^4N_\mathrm{c}^2I_\ell^\qty(4/3)\omega^{4/3}\bigg],
    \end{split}
    \end{align}
    with another family of integrals $I_\ell^\qty(q)$, given by
    \begin{align}
    \begin{split}
        \label{eq:C_l-cs-integrals}
        I_\ell^{(q)}\qty(\omega)&\equiv2\uppi\int_0^\uppi\dd{\theta_\mathrm{o}}\sin\theta_\mathrm{o} P_\ell\qty(\cos\theta_\mathrm{o})I_C^{(q)}\qty(\theta_\mathrm{o},\omega)\\
        &=2\uppi\int_0^{+\infty}\dd{u}uP_\ell\qty(\frac{1-u^2}{1+u^2})I_C^{(q)}\qty(\theta_\mathrm{o},\omega)
    \end{split}
    \end{align}

The expression Eq.~\eqref{eq:C_l-cs} with the integrals Eq.~\eqref{eq:C_l-cs-integrals} and the limits Eq.~\eqref{eq:2pcf-u-limits} allows us to calculate the $C_\ell$ components describing the anisotropy in the SGWB, sourced by cosmic strings with any given loop distribution, accurate for
    \begin{equation}
        \ell\lesssim\ell_\mathrm{max}\equiv\uppi/\theta_\mathrm{min}.
    \end{equation}

\subsection{Generalized loop distribution}
\begin{figure*}[t]
    \includegraphics[width=\textwidth]{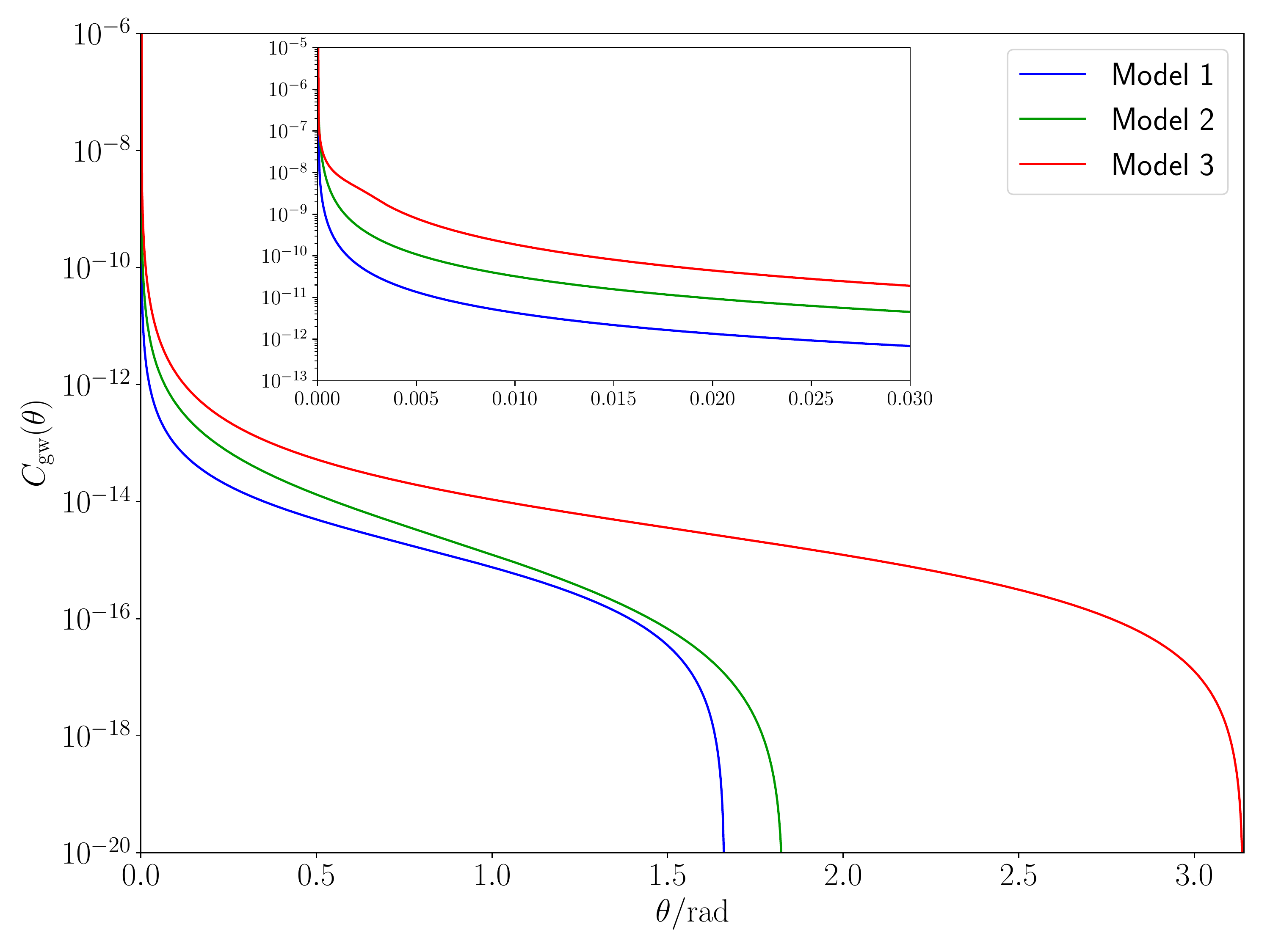}
    \caption{The angular dependence of the 2PCF $C_\mathrm{gw}\qty(\theta_\mathrm{o},\nu_\mathrm{o})$ in each of the three models, at frequency $\nu_\mathrm{o}=10^{-8}\mathrm{Hz}$, and with $G\mu=10^{-7}$, $N_\mathrm{c}=N_\mathrm{k}=1$, and $k=1$. The subplot shows the behaviour for small angles $\theta_\mathrm{o}\lesssim1^\circ$.}
    \label{fig:2pcf-models}
\end{figure*}
\begin{figure*}[t]
    \includegraphics[width=\textwidth]{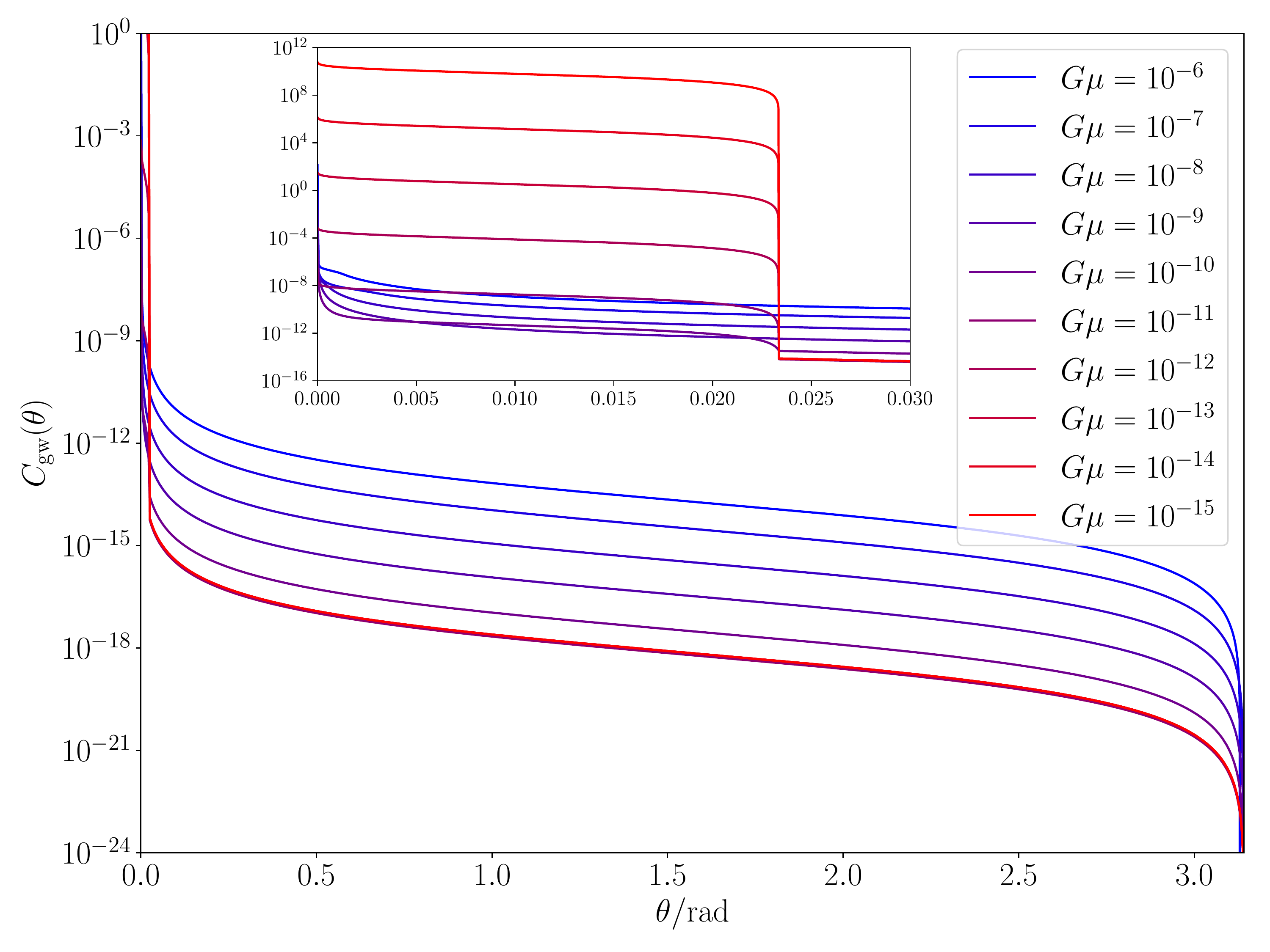}
    \caption{The angular dependence of the 2PCF $C_\mathrm{gw}\qty(\theta_\mathrm{o},\nu_\mathrm{o})$ in model 3 for a range of values of $G\mu$, at frequency $\nu_\mathrm{o}=10^{-8}\mathrm{Hz}$, and with $N_\mathrm{c}=N_\mathrm{k}=1$, and $k=1$. The subplot shows the behaviour for small angles $\theta_\mathrm{o}\lesssim1^\circ$.}
    \label{fig:2pcf-Gmu}
\end{figure*}
Following Ref.~\cite{Abbott:2017mem} we will consider three distinct analytic models of cosmic string loop distributions, with the common property that the string dynamics are obtained through the Nambu-Goto action and that string intercommutation occurs with unit probability.
As in Ref.~\cite{Abbott:2017mem}, we call these models $M=1,2,3$, defined as follows.
    \begin{itemize}
        \item Model $M=1$: This assumes that in the scaling regime, all loops chopped off the superhorizon string network are formed with the same relative size~\cite{Vilenkin:2000jqa}, which we will denote by $\alpha$ with a subscript ``r" or ``m" indicating whether we refer to the radiation or the matter era, respectively.
        \item Model $M=2$: Extrapolating from the loop production function found in numerical simulations, this analytic model~\cite{Blanco-Pillado:2013qja} gives the distribution of string loops of given size at fixed time, under the assumption that the momentum dependence of the loop production function is weak.
        \item Model $M=3$: Using a numerical simulation~\cite{Ringeval:2005kr}---different from the one leading to model $M=2$---this analytic model~\cite{Lorenz:2010sm} gives the distribution of non-self intersecting loops at a given time. This model includes a new length scale, the \emph{gravitational back-reaction scale}, $\gamma_{\mathrm{c}}$, with $\gamma_{\mathrm{c}} <\gamma_{\mathrm{d}}$, leading to a different loop distribution than model 2 for the smallest loops.
    \end{itemize}
We give below the general expression for the loop distribution in the radiation and matter eras, and specify the values of the parameters for each of the three models defined above.
Radiation era:
    \begin{align}
    \begin{split}
        \label{eq:radiation-era-distribution}
        \bar{\mathcal{F}}_\mathrm{r}&\approx C_\mathrm{r}\gamma^{-p_\mathrm{r}-1}\Theta\qty(\alpha_\mathrm{r}-\gamma)\Theta\qty(\gamma-\gamma_\mathrm{d})\\
        &+C_\mathrm{r}\qty(1-\frac{3}{2p_\mathrm{r}})^{p_C}\gamma_\mathrm{d}^{-1}\gamma^{-p_\mathrm{r}}\Theta\qty(\gamma_\mathrm{d}-\gamma)\Theta\qty(\gamma-\gamma_\mathrm{cr})\\
        &+C_\mathrm{r}\qty(1-\frac{3}{2p_\mathrm{r}})^{p_C}\gamma_\mathrm{d}^{-1}\gamma_\mathrm{cr}^{-p_\mathrm{r}}\Theta\qty(\gamma_\mathrm{cr}-\gamma).
    \end{split}
    \end{align}
Matter era:
    \begin{align}
    \begin{split}
        \label{eq:matter-era-distribution}
        \bar{\mathcal{F}}_\mathrm{m}&\approx \qty(C_\mathrm{m1}-C_n\gamma^{0.31})\gamma^{-p_{\mathrm{m}1}-1}\Theta\qty(\gamma-\beta)\\
        &\qquad\qquad\qquad\qquad\qquad\qquad\times\Theta\qty(\alpha_\mathrm{m}-\alpha_\tau\frac{a_\mathrm{eq}^{1/2}}{\tau^{1/3}}-\gamma)\\
        &+C_{\mathrm{m}2}\qty(\frac{a_\mathrm{eq}^{3/4}}{\tau^{1/2}})^{p_\tau}\gamma^{-p_{\mathrm{m}2}-1}\Theta\qty(\beta-\gamma)\Theta\qty(\gamma-\gamma_\mathrm{d})\\
        &+C_{\mathrm{m}2}\qty(1-\frac{1}{p_\mathrm{m1}})^{p_C}\gamma_\mathrm{d}^{-1}\gamma^{-p_{\mathrm{m}2}}\Theta\qty(\gamma_\mathrm{d}-\gamma)\Theta\qty(\gamma-\gamma_\mathrm{cm})\\
        &+C_{\mathrm{m}2}\qty(1-\frac{1}{p_\mathrm{m1}})^{p_C}\qty(\frac{a_\mathrm{eq}^{3/4}}{\tau^{1/2}})^{p_\tau}\gamma_\mathrm{d}^{-1}\gamma_\mathrm{cm}^{-p_{\mathrm{m}2}}\Theta\qty(\gamma_\mathrm{cm}-\gamma).
    \end{split}
    \end{align}
Model 1:
    \begin{align}
    \begin{split}
        C_\mathrm{r}&=C_\mathrm{m2}\approx1.6,\quad C_\mathrm{m1}\approx0.48,\quad C_\mathrm{n}=0,\\
        p_\mathrm{r}&=p_\mathrm{m2}=\frac{3}{2},\quad p_\mathrm{m1}=1,\quad p_C=0,\quad p_\tau=1,\\
        \gamma_\mathrm{cr}&=\gamma_\mathrm{cm}=\gamma_\mathrm{d},\quad\alpha_\mathrm{r}=\alpha_\mathrm{m}\approx0.1,\quad\alpha_\tau=0.
    \end{split}
    \end{align}
Model 2:
    \begin{align}
    \begin{split}
        C_\mathrm{r}&=C_\mathrm{m2}\approx0.18,\quad C_\mathrm{m1}\approx0.27,\quad C_\mathrm{n}\approx0.45,\\
        p_\mathrm{r}&=p_\mathrm{m2}=\frac{3}{2},\quad p_\mathrm{m1}=1,\quad p_C=0,\quad p_\tau=1,\\
        \gamma_\mathrm{cr}&=\gamma_\mathrm{cm}=\gamma_\mathrm{d},\quad\alpha_\mathrm{r}\approx0.1,\quad\alpha_\mathrm{m}\approx0.18,\quad\alpha_\tau=0.
    \end{split}
    \end{align}
Model 3:
    \begin{align}
    \begin{split}
        C_\mathrm{r1}&\approx0.08,\quad C_\mathrm{m1}=C_\mathrm{m2}\approx0.016,\quad  C_\mathrm{n}=0,\\
        p_\mathrm{r}&\approx1.60,\quad p_\mathrm{m1}=p_\mathrm{m2}\approx1.41,\quad p_C=1,\quad p_\tau=0,\\
        \gamma_\mathrm{cr}&\approx20\qty(G\mu)^{3-p_\mathrm{r}},\quad\gamma_\mathrm{cm}\approx20\qty(G\mu)^{3-p_\mathrm{m1}},\\
        \alpha_\mathrm{r}&=2,\quad\alpha_\mathrm{m}=3,\quad\alpha_\tau=1.
    \end{split}
    \end{align}
In all models,
    \begin{align}
    \begin{split}
        \beta\qty(\tau)&\equiv\frac{a_\mathrm{eq}^{3/2}}{\tau}\qty(\alpha_\mathrm{r}+\gamma_\mathrm{d})-\gamma_\mathrm{d}\approx\frac{a_\mathrm{eq}^{3/2}\alpha_\mathrm{r}}{\tau},\\
        \gamma_\mathrm{d}&=\Gamma G\mu\approx50G\mu.
    \end{split}
    \end{align}
We have used the approximation
    \begin{equation}
        \qty(\gamma+\gamma_\mathrm{d})^n\approx
        \left\{
        \begin{matrix*}[l]
            \gamma_\mathrm{d}^n,&0\le\gamma<\gamma_\mathrm{d},\\
            \gamma^n,&\gamma_\mathrm{d}\le\gamma\le\gamma^*,
        \end{matrix*}
        \right.
    \end{equation}
    which is very accurate for $\gamma\gg\gamma_\mathrm{d}$ and $\gamma\ll\gamma_\mathrm{d}$, and is correct to within an order of magnitude around $\gamma\approx\gamma_\mathrm{d}$.

\subsection{Results and discussion}
\label{section-results-and-discussion}
\begin{figure*}[t]
    \includegraphics[width=\textwidth]{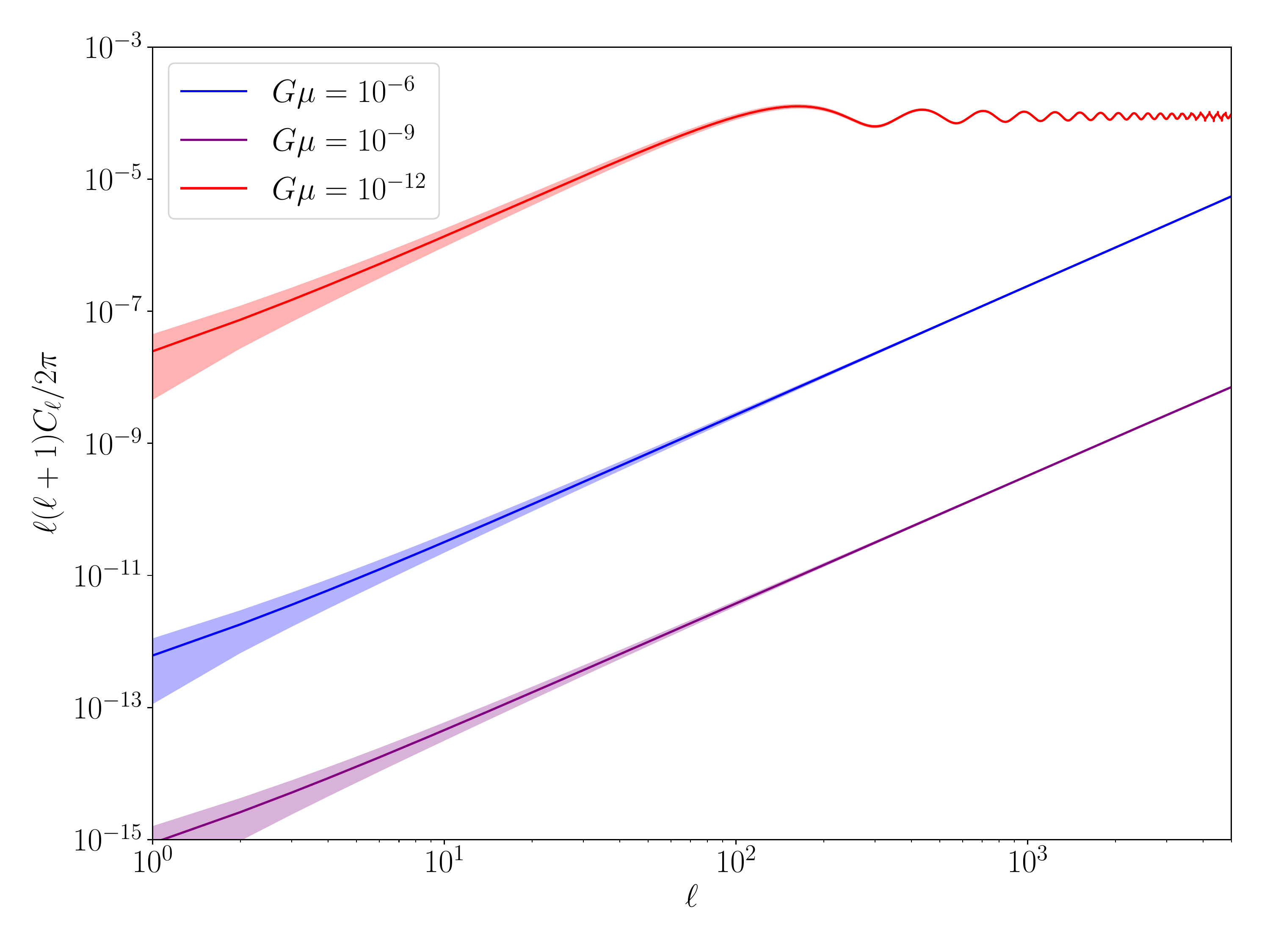}
    \caption{The approximate contribution to the anisotropic variance of $\Omega_\mathrm{gw}$ as a function of $\ln\ell$ in model 3, shown for three values of $G\mu$, with $\nu_\mathrm{o}=10^{-8}\mathrm{Hz}$, $N_\mathrm{c}=N_\mathrm{k}=1$, and $k=1$. Uncertainty in the $C_\ell$'s due to cosmic variance is shown by the shaded regions.}
    \label{fig:log-var}
\end{figure*}
We have evaluated the integrals in Eqs.~\eqref{eq:monopole-integrals}, \eqref{eq:dipole-integrals}, \eqref{eq:lambda-integrals} and \eqref{eq:2pcf-u-integrals} for the generalized loop distribution Eqs.~\eqref{eq:radiation-era-distribution} and \eqref{eq:matter-era-distribution} to find expressions for the monopole, kinematic dipole, and 2PCF of the corresponding loop network.
These expressions are very lengthy, and are not reproduced here.
It is worth stressing that this process is almost entirely analytical, minimizing the computational cost involved.
The only numerical elements of our analysis are a root-finding process used to calculate $\tau_*$ from Eq.~\eqref{eq:duty-cycle} and an ensemble of numerical integrations over the 2PCF to give the $C_\ell$ spectrum.

Figure \ref{fig:monopole-models} shows the monopole for each of the three models at a fixed $G\mu$, while Fig.~\ref{fig:monopole-Gmu} shows how the monopole depends on $G\mu$ for model 3.
It is interesting to note how, as well as the obvious overall suppression in $\bar{\Omega}_\mathrm{gw}$ for smaller $G\mu$, the spectrum is also pushed to higher frequencies when $G\mu$ is decreased.
Physically, this is because decreasing $G\mu$ decreases the typical size of the loops, and therefore \emph{increases} the lower bound on the emitted frequency, which goes as $1/\gamma t$ (we refer the reader to Sec.~\ref{sec:cusps-kinks-collisions}).

Figure \ref{fig:2pcf-models} shows the 2PCF for each of the three models.
For all of the 2PCF plots shown, we have selected a frequency of $10^{-8}$~Hz to illustrate our results, partly because this is near the peak value of $\bar{\Omega}_\mathrm{gw}$ for the larger values of $G\mu$ we consider, and partly because it lies within the frequency range of pulsar timing arrays.
We can see that for large angles, the correlation is many orders of magnitude smaller in models 1 and 2 than in model 3.
This is because the anisotropy at large angular scales is related to the largest loops in the network; in models 1 and 2, the loops are limited to $\gamma\lesssim0.1$, while model 3 allows loops to be an order of magnitude larger than this.
For smaller angles, the angular dependence of the correlation is essentially the same for all three models, with the only apparent difference being an overall constant factor.
Since the correlation is so much stronger on this angular scale, it is this regime that will govern the observable anisotropies.
As the three models are so similar in terms of the angular dependence of the 2PCF on this scale, the rest of our results focus on model 3, which has the strongest correlation, and therefore the most prominent anisotropies.

Since we are considering small angular scales, it is important to check that we are above the scale set by $\theta_\mathrm{min}$ [given by Eq.~\ref{eq:theta-min}], as our expressions may be inaccurate for angles smaller than this (for reasons outlined in Sec.~\ref{sec:2pcf}).
We find that in the cases we consider $\theta_\mathrm{min}$ is always less than $10^{-5}$~rad (roughly one arcsecond), and thus does not pose a problem for the models explored here.

We also use the condition Eq.~\eqref{eq:grf-condition} to check that the cosmic string SGWB is Gaussian.
We find that the duty cycle for all sources included in the background is extremely large, typically on the order of $10^{30}$.
This means that even though the maps shown in Figs.~\ref{fig:maps} and~\ref{fig:maps-zoom} were produced for a very low-frequency regime of the SGWB and with a very high angular resolution (both of which generally make Gaussianity harder to achieve), we have $\qty(\frac{N_\text{pix}}{\Lambda}+\frac{N^2_\text{pix}}{\Lambda^2})/\nu_\text{o}\approx10^{-14}\text{ s}$, and the background is a GRF to an extremely good approximation, even after an extremely short observing time.
\begin{figure*}[t]
    \includegraphics[width=\textwidth]{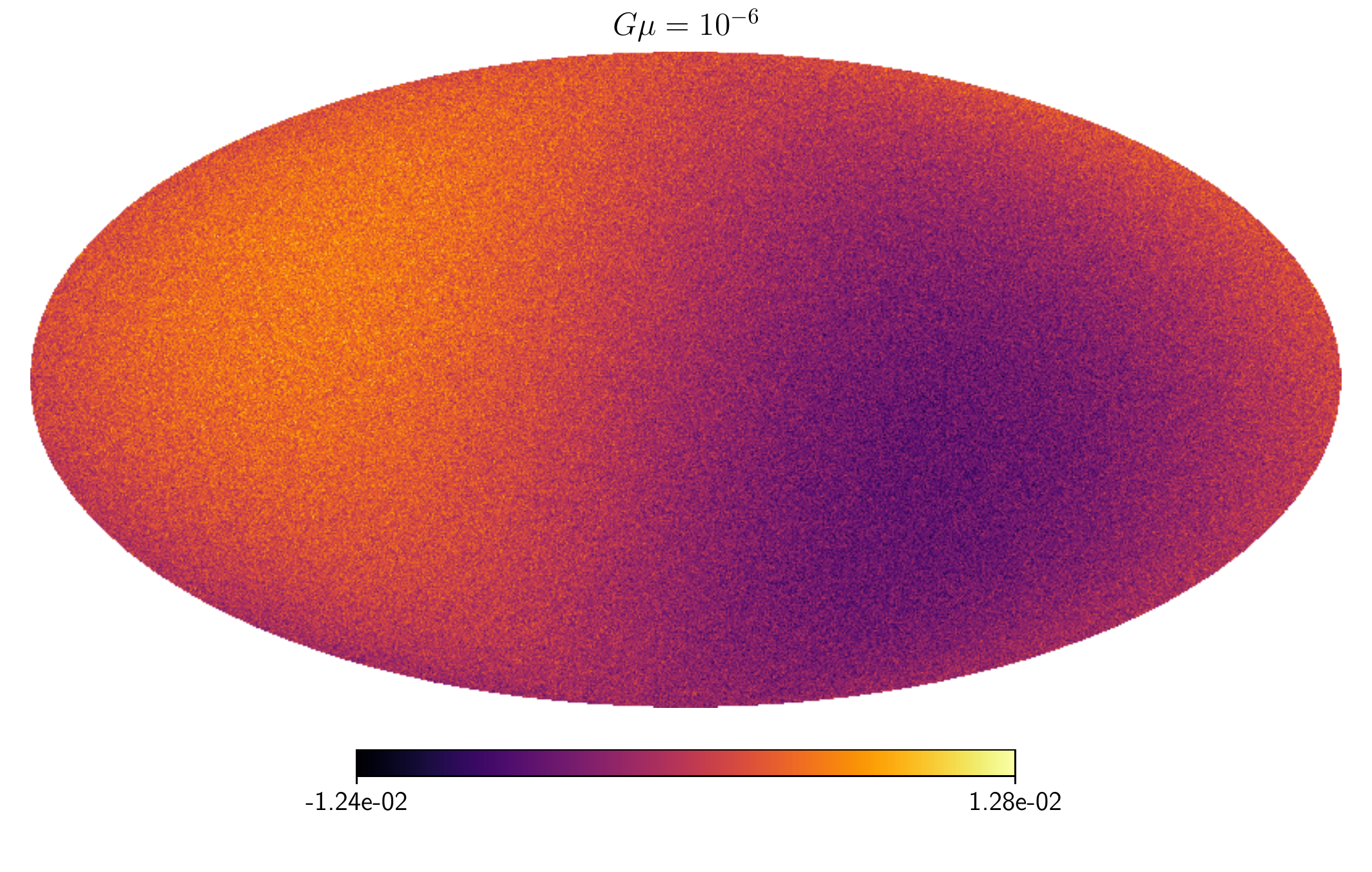}
    \caption{A random realization of the SGWB using the first 5,000 $\ell$-modes and \emph{including the kinematic dipole} for $G\mu=10^{-6}$, using model 3 with $\nu_\mathrm{o}=10^{-8}\mathrm{Hz}$, $N_\mathrm{c}=N_\mathrm{k}=1$, and $k=1$. This is generated with an angular resolution of $\approx50$ arcseconds.}
    \label{fig:map-dipole}
\end{figure*}

Figure \ref{fig:2pcf-Gmu} shows how the 2PCF depends on $G\mu$.
Recall that $C_\mathrm{gw}$ is normalized with respect to the isotropic energy density such that only the \emph{relative} amplitude of the anisotropies matters---while the \emph{absolute} energy density decreases with $G\mu$, that will not be reflected here.
We can see that at large angles, the correlation decreases by roughly an order of magnitude for each order of magnitude decrease in $G\mu$, until around $G\mu\approx10^{-11}$ where the correlation seems to reach a minimum as a function of $G\mu$.
For smaller angles $\theta_\mathrm{o}\lesssim1^\circ$, we see that decreasing $G\mu$ causes the correlation to decrease gradually, until $G\mu$ goes below around $10^{-11}$, which causes an exponential increase in the correlation at small angles.
Physically, we expect that this is due to a trade-off between two effects associated with a decrease in $G\mu$: fewer signals, and less energy density per signal.
Reducing the energy per signal will mean that the typical amplitude of the anisotropies will decrease, explaining the initial drop in the correlation as $G\mu$ is decreased from $10^{-6}$ to around $10^{-11}$.
However, for small enough $G\mu$ the dominant effect is the suppression of the number of signals, which makes the SGWB more discretized and therefore increases its angular granularity to such an extent that the anisotropic fluctuations become much larger (see also Figs.~\ref{fig:maps} and~\ref{fig:maps-zoom}).
This has important ramifications for the detectability of a cosmic string network, as the relatively more prominent anisotropies produced by a smaller $G\mu$ could plausibly increase the detection prospects for a SGWB that would otherwise be too faint.

It is interesting to note that the small-angle enhancement in the correlation suddenly ``switches on" for angles less than $\approx0.023$~rad.
In fact, this is the angle $\theta_\mathrm{eq}$ defined in Eq.~\eqref{eq:theta-eq}, corresponding to the maximum angular size of features in the radiation era.
We therefore see in Fig.~\ref{fig:2pcf-Gmu} that, when $G\mu$ is sufficiently small, there is a much stronger correlation for bursts originating in the radiation era.
The reason for the abruptness of this transition as we vary $\theta_\mathrm{o}$ is simply due to our lack of a smooth transition between the matter and radiation eras.
However, we feel that our results capture the most important features of $C_\mathrm{gw}$, and that implementing a smooth transition will not change our results too drastically.

Figure \ref{fig:log-var} plots $\ell(\ell+1)C_\ell/2\uppi$ against $\ell$.
As mentioned in Sec.~\ref{sec:formalism-sgwb-decomposition}, this can roughly be thought of as the contribution to the total variance in $\Omega_\mathrm{gw}$ per logarithmic bin in $\ell$.
As we would expect from the results in Fig.~\ref{fig:2pcf-Gmu}, the variance decreases as we go from $G\mu=10^{-6}$ to $G\mu=10^{-9}$ due to a reduction in the energy per signal, but then increases greatly as we go from $G\mu=10^{-9}$ to $G\mu=10^{-12}$ due to the increased granularity of the SGWB.
We see that, regardless of the value of $G\mu$, this quantity increases exponentially as we go to higher $\ell$-modes, meaning that the anisotropies are characterized by small angular scales in every case.
We also see that for large enough $\ell$, this variance contribution eventually reaches a plateau (with small oscillations)---this is not shown explicitly for $G\mu=10^{-6},10^{-9}$, but will occur at very high $\ell$-modes, $\ell\gtrsim$~10,000 or so.
This is to be expected, as it ensures that the total variance is finite.

Figure \ref{fig:maps} shows random realizations of the SGWB, created from our $C_\ell$ coefficients (up to $\ell=$~5,000) using the HEALPix package.\footnote{ \url{http://healpix.sourceforge.net}}
The angular features are small and somewhat difficult to discern, so we show in Fig.~\ref{fig:maps-zoom} a $10^\circ\times10^\circ$ portion of each map, magnifying the angular fluctuations.
As expected given our other results, the angular features appear much larger and more distinct for $G\mu=10^{-12}$ than the other cases, though we remind the reader that the \emph{absolute} values for the energy density are much smaller than in the other cases.
While these maps are useful for illustrative purposes, we emphasize the main physical content of our results is in the 2PCF (as shown in Fig.~\ref{fig:2pcf-Gmu}).
The maps are just convenient visualizations of the angular correlation.

Figure \ref{fig:map-dipole} shows one of these maps with the kinematic dipole included (using the simple result from the Appendix), to illustrate how it obscures the small-scale anisotropies.
This shows how important it is from an observational point of view to be able to remove this dipole.

\subsection{A note on previous works}
\begin{figure*}[p]
    \begin{center}
        \includegraphics[height=0.3\textheight]{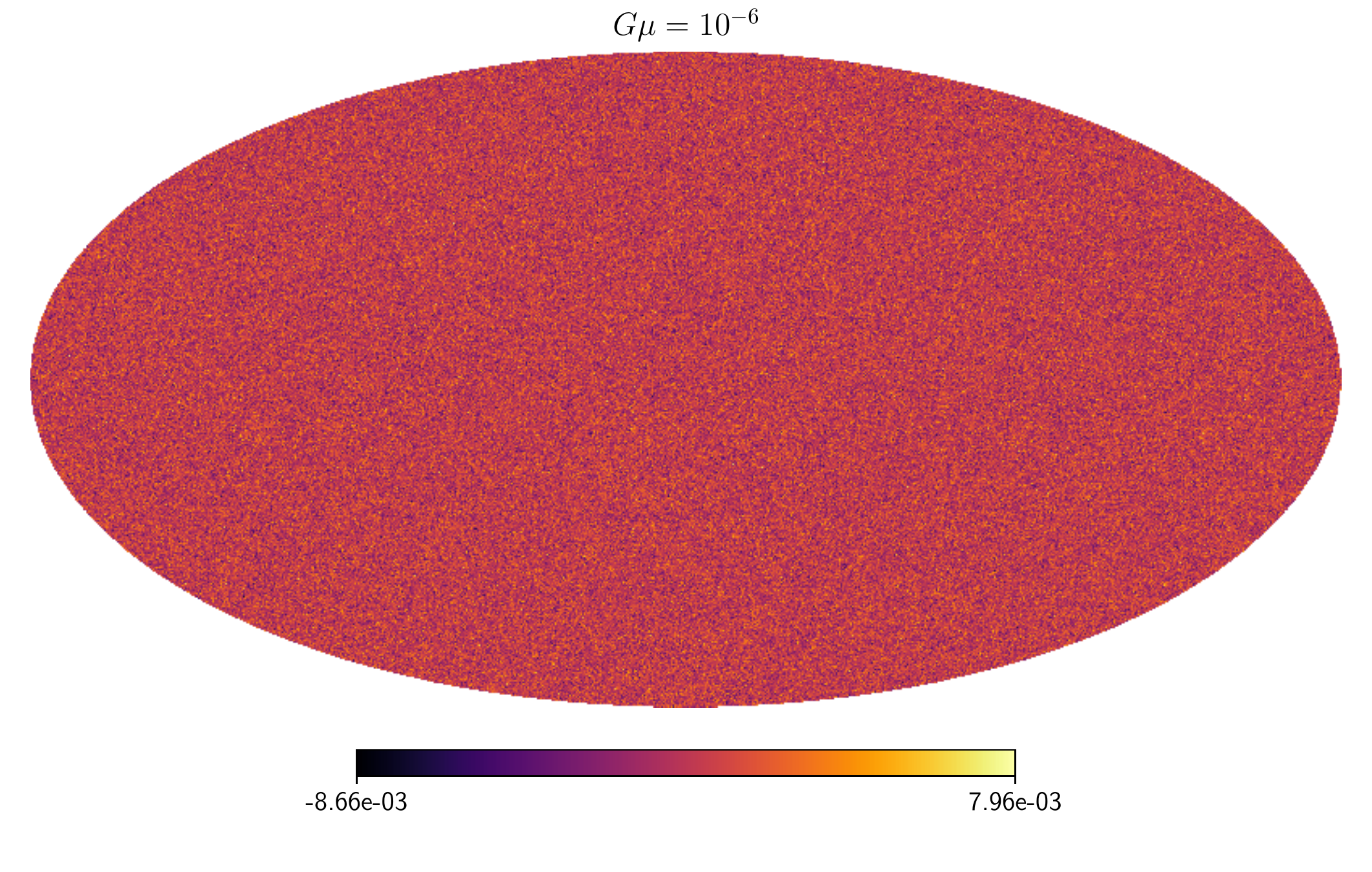}
        \includegraphics[height=0.3\textheight]{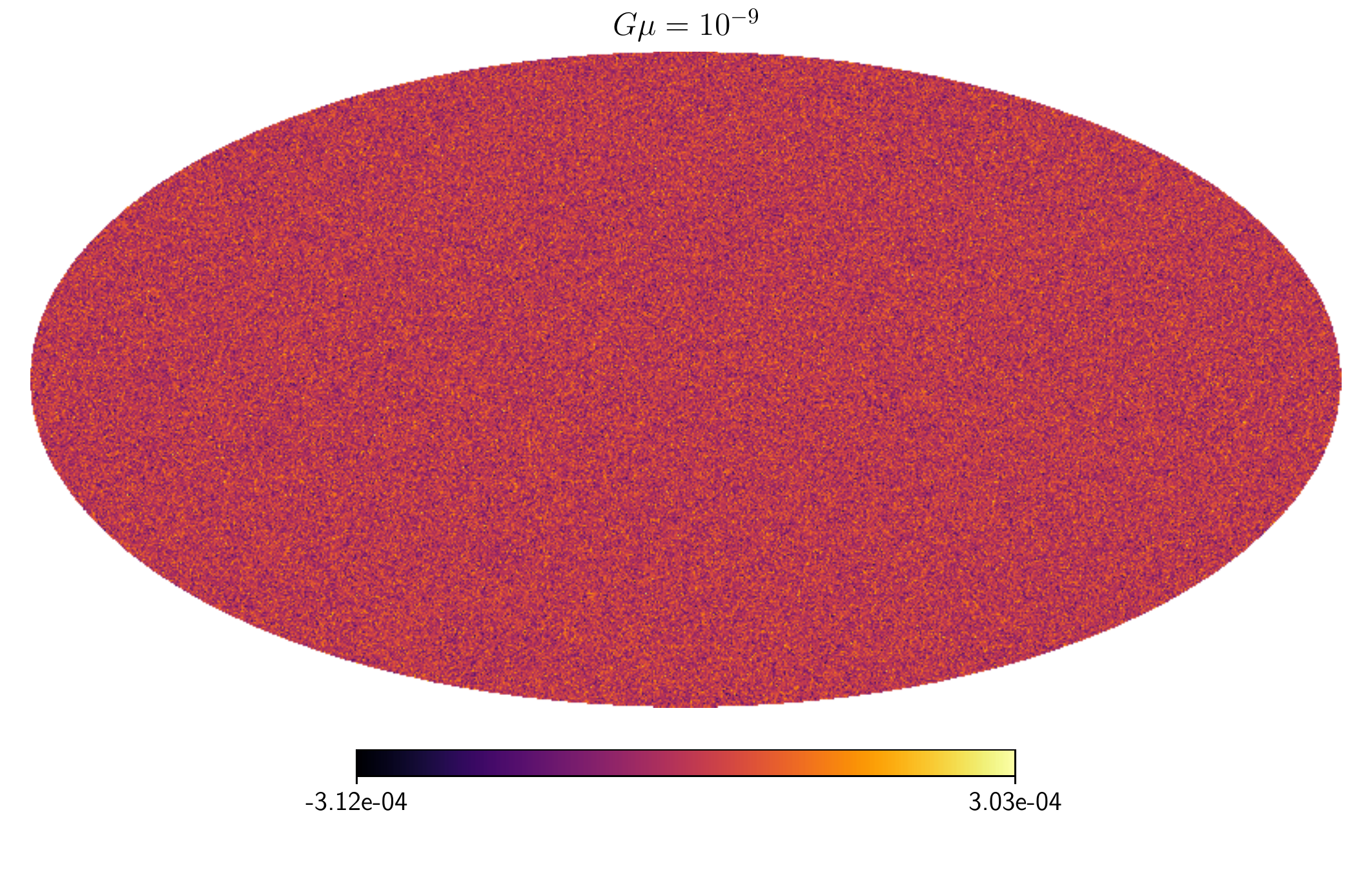}
        \includegraphics[height=0.3\textheight]{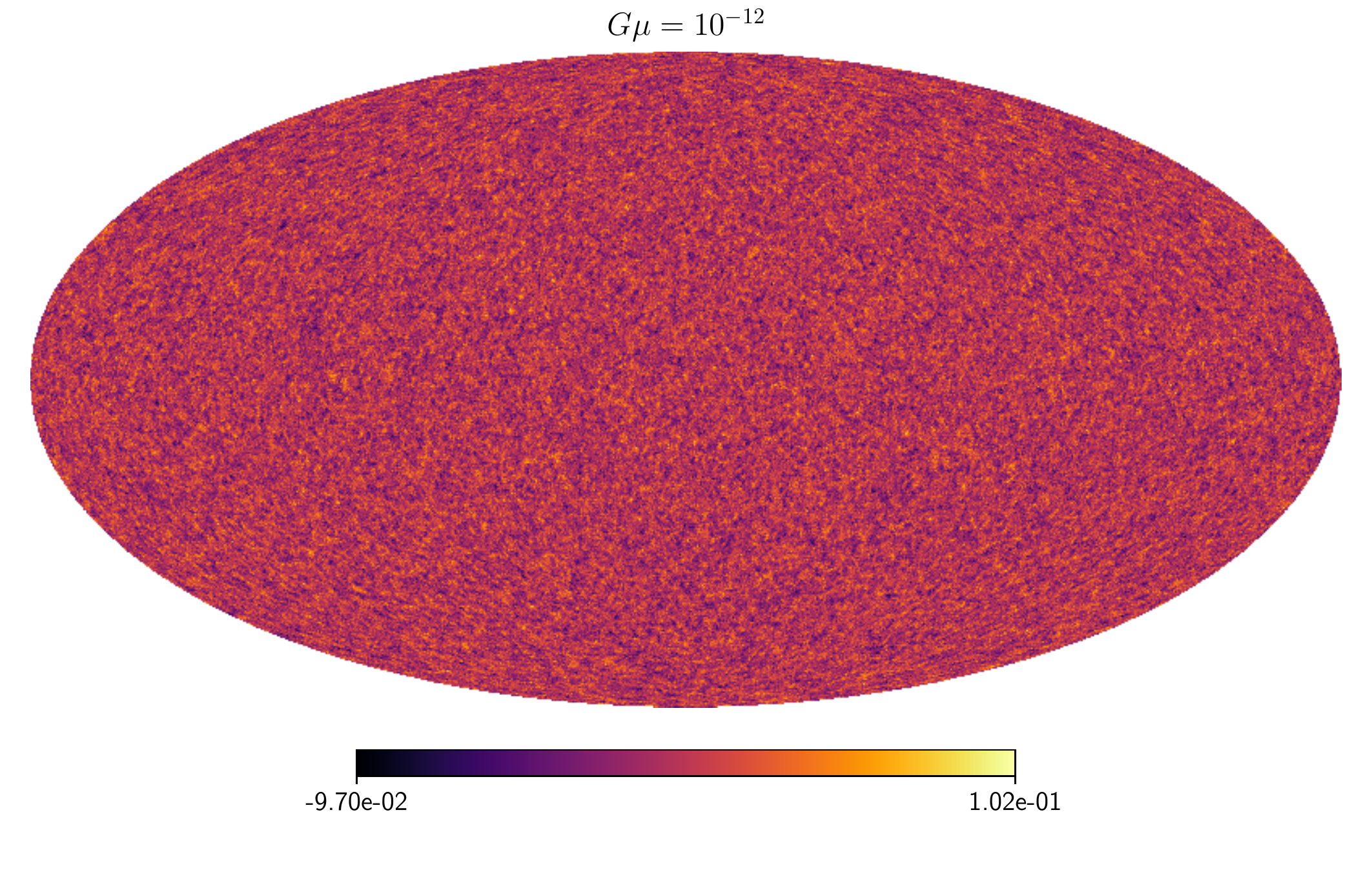}
    \end{center}
    \caption{Random realizations of the SGWB using the first 5,000 $\ell$-modes for three values of $G\mu$, using model 3 with $\nu_\mathrm{o}=10^{-8}\mathrm{Hz}$, $N_\mathrm{c}=N_\mathrm{k}=1$, and $k=1$. These are generated with an angular resolution of $\approx50$ arcseconds.}
    \label{fig:maps}
\end{figure*}
\begin{figure*}[tp]
    \begin{center}
        \includegraphics[height=0.45\textheight]{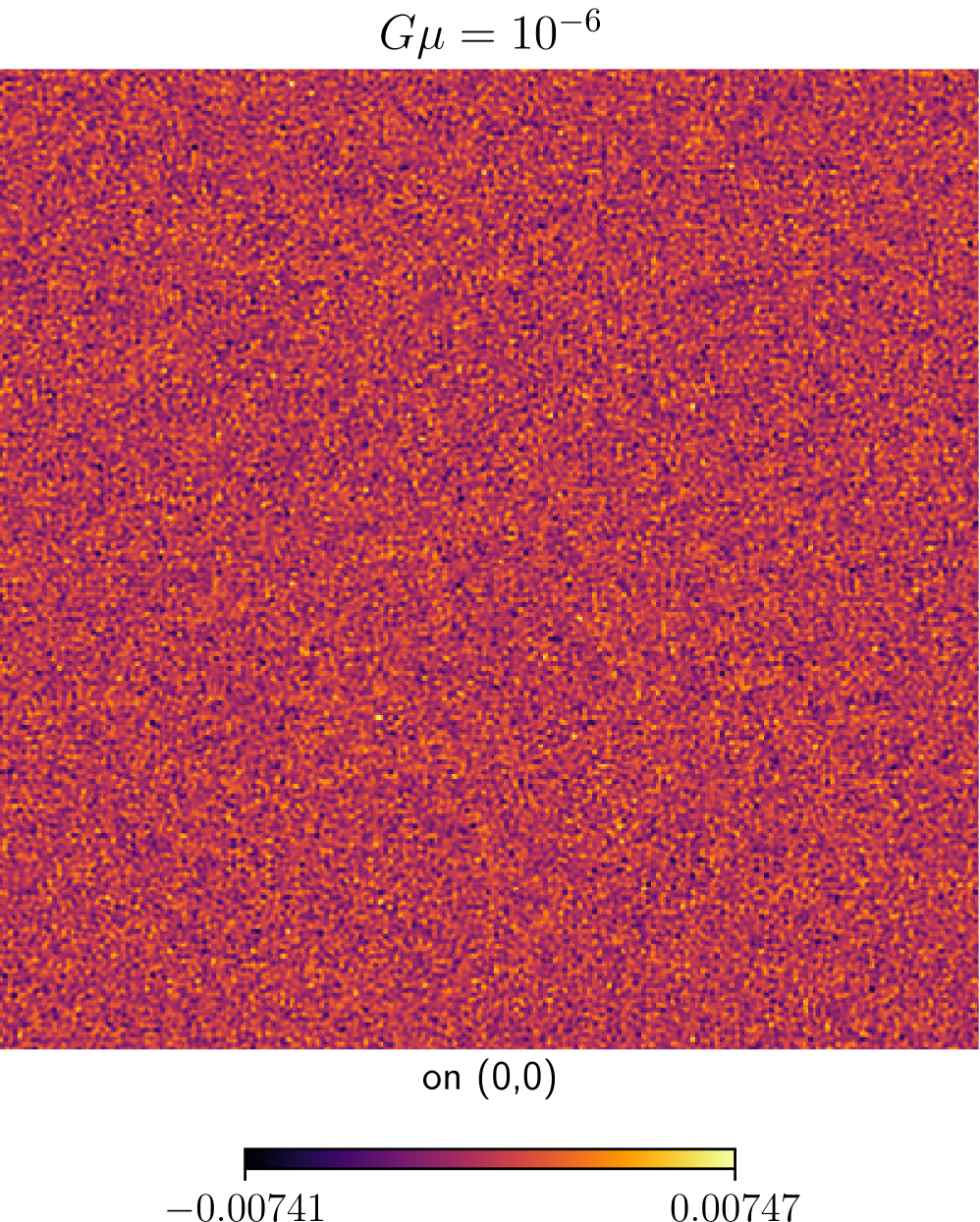}
        \includegraphics[height=0.45\textheight]{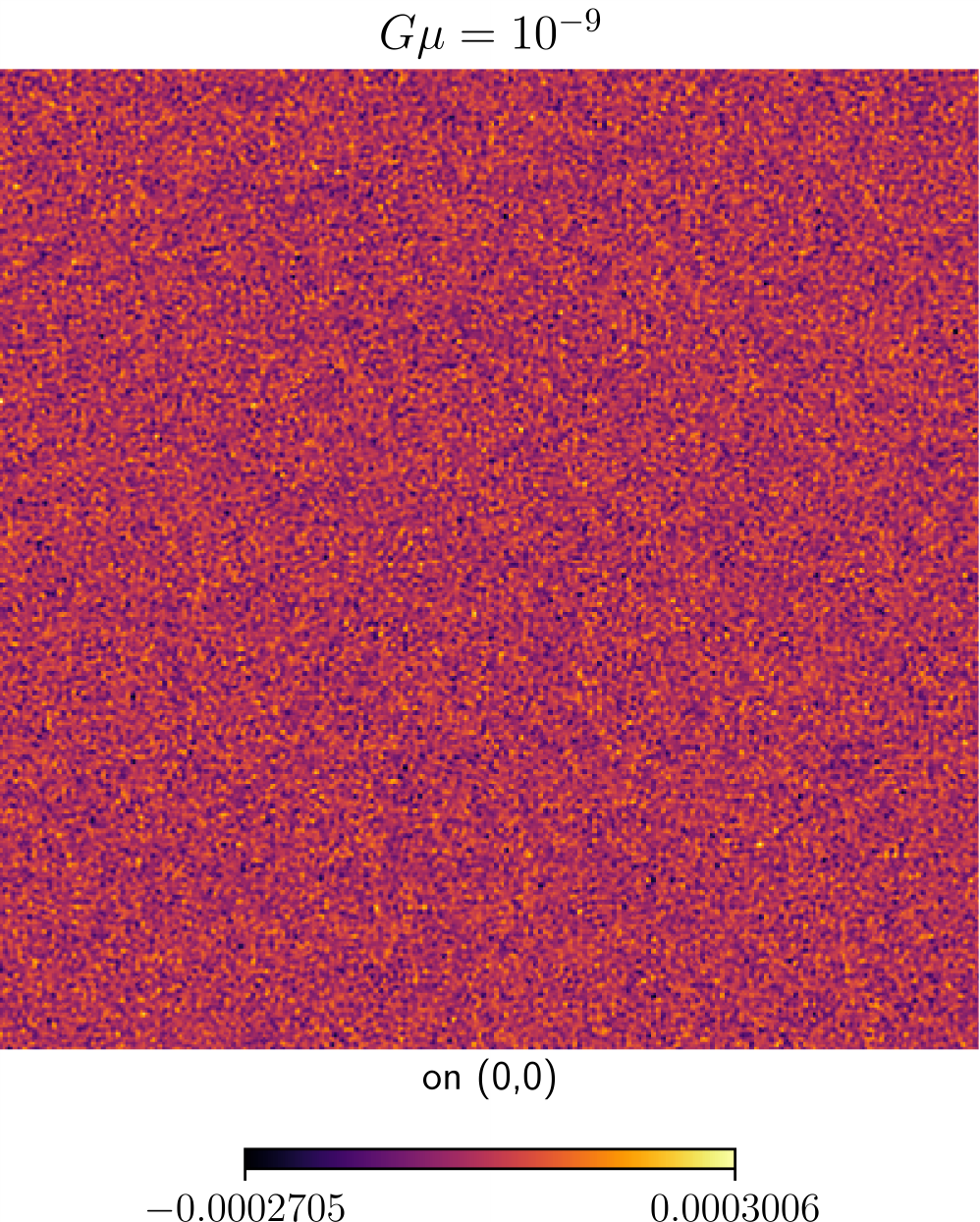}
        \includegraphics[height=0.45\textheight]{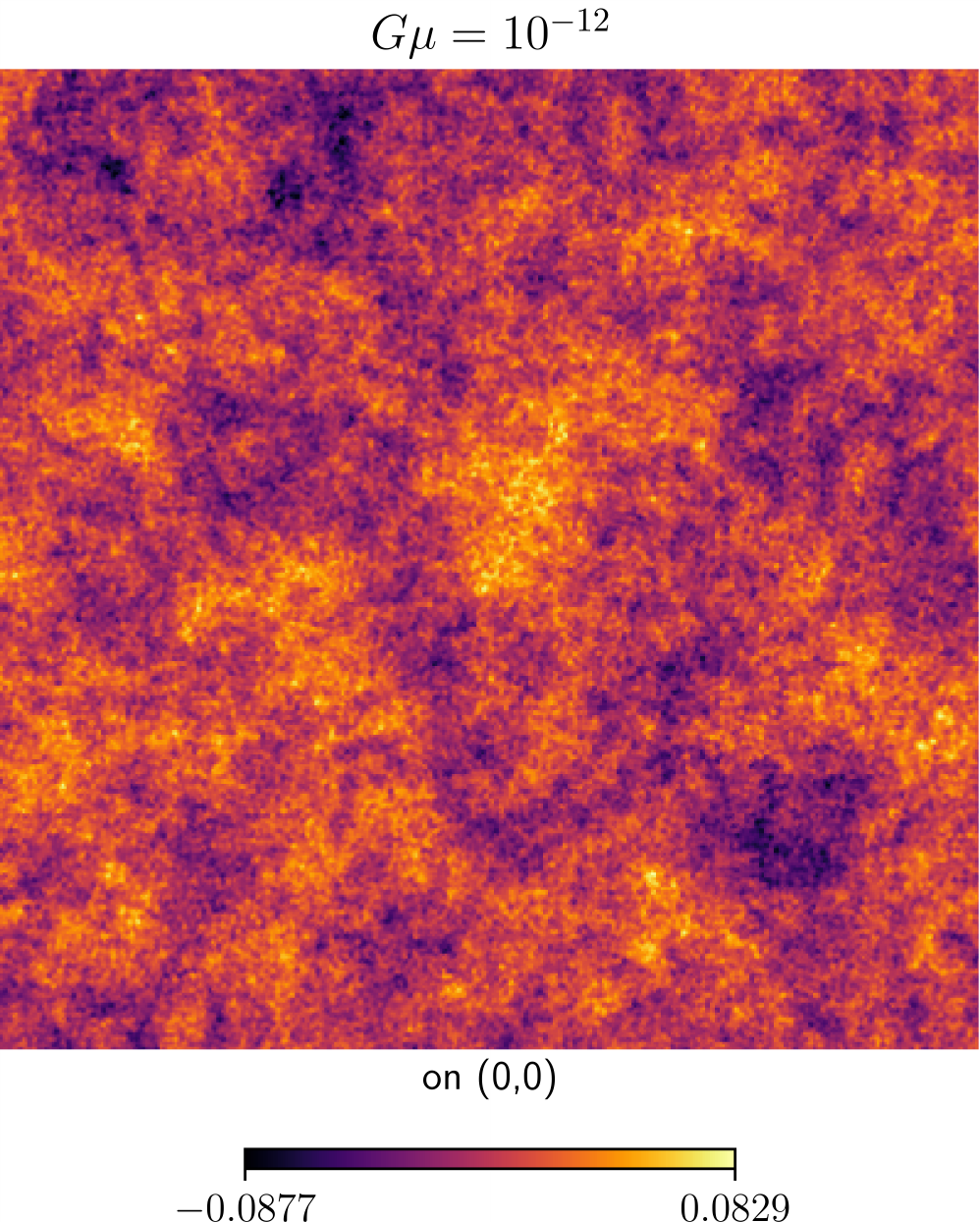}
    \end{center}
    \caption{Magnified $10^\circ\times10^\circ$ regions of the maps shown in Fig.~\ref{fig:maps}. These are generated with an angular resolution of $\approx50$ arcseconds.}
    \label{fig:maps-zoom}
\end{figure*}
We mention briefly two previous articles (Refs.~\cite{Kuroyanagi:2016ugi} and \cite{Olmez:2011cg}) which considered SGWB anisotropies from cosmic strings, and discuss how this work differs from them.

First, we note that the sources in Ref.~\cite{Kuroyanagi:2016ugi} are randomly distributed on the sky, i.e. they have \emph{no spatial correlation}.
It is for this reason that the correlation function $C_\mathrm{gw}$ is not considered in that work.
The $C_\ell$'s are instead coefficients in a multipole expansion of $\Omega_\mathrm{gw}$, and no comparison should be made between them and the $C_\ell$'s calculated in this work.

Second, while Ref.~\cite{Olmez:2011cg} does consider the 2PCF, the loop network model used therein is a modified version of what we call model 1, but with extremely small loops ($\alpha$ is replaced by $\epsilon\alpha$, where $\epsilon=10^{-11}$ in the case shown in their Fig.~1).
This is very different from any of the models we have considered here.

\section{Conclusion}
We have developed a powerful formalism for producing analytical predictions of the frequency spectrum and angular correlation of the (anisotropic) SGWB, applicable to any astrophysical or cosmological source.
This builds upon the results of previous works (in particular Ref.~\cite{Cusin:2017fwz}) in a number of ways.
First, the directional SGWB energy density parameter $\Omega_\mathrm{gw}$ is written explicitly in terms of the strain spectrum $\tilde{h}$ of the source---this was not the case in Ref.~\cite{Cusin:2017fwz}, and doing so makes the application of the formulae more straightforward.
Second, we derive a sufficient condition for the SGWB to be a Gaussian random field, thereby justifying the use of the 2PCF; we find that it is not necessary for the duty cycle to be large (as is often assumed for ``Gaussian backgrounds" in the literature), and that any GW background can in principle be made Gaussian by increasing the observing time $T$ (with the caveat that this not the case for the background from compact binary coalescences---see Ref.~\cite{Jenkins:2018uac}).
Third, we use the duty cycle as a function of distance to carefully distinguish between foreground and background signals, and thereby isolate the Gaussian part of the SGWB, which is desirable for the study of the anisotropies.
We also give an expression for the expected magnitude of the kinematic dipole, which will enable us to isolate the cosmological anisotropies from any observations (as can be seen in Fig.~\ref{fig:map-dipole}, failing to remove this dipole interferes significantly with the angular statistics of the SGWB).
Finally, we discuss how to relate our analytical predictions for the 2PCF to observed quantities, taking into account cosmic variance.

We have applied this formalism to the case of cosmic strings (specifically, Nambu-Goto string loop networks).
The most interesting results are that the angular spectrum of the 2PCF is relatively insensitive to our choice of model, differing only by a constant factor at small scales, and that decreasing the value of $G\mu$ can produce much stronger relative anisotropies.
These anisotropies are characterized by small angular scales ($\theta_\mathrm{o}\lesssim1^\circ$), and are primarily due to radiation-era sources.
Our results have interesting implications for the prospects of detecting cosmic strings, and may be exploited in future observational work.

The formalism in Sec.~\ref{sec:general-formalism} is not limited to cosmic strings, and we plan to apply it to a variety of GW sources.
This includes a study of the astrophysical background from compact binaries in Ref.~\cite{Jenkins:2018uac}.

\begin{acknowledgments}
Many thanks to Joe Romano for questions and discussion that prompted us to include Sec.~\ref{sec:observations}, and to Tania Regimbau for useful comments on Sec.~\ref{sec:gaussian-non-gaussian}.
Figures~\ref{fig:map-dipole}, \ref{fig:maps}, and \ref{fig:maps-zoom} were created using the HEALPix package~\cite{Gorski:2004by}.
A.C.J. is supported by King's College London through a Graduate Teaching Scholarship.
M.S. is supported in part by the Science and Technology Facility Council (STFC), UK, under the research grant ST/P000258/1.
\end{acknowledgments}

\appendix*
\section{Including the kinematic dipole in the correlation function}
We have decomposed the SGWB energy density contrast $\delta_\mathrm{gw}$ into a term associated with the sources and a term encoding the kinematic dipole, $\delta_\mathrm{gw}\equiv\delta^{(\mathrm{s})}_\mathrm{gw}+\mathcal{D}\vu*e_\mathrm{o}\vdot\vu*v_\mathrm{o}$, and have defined $C_\mathrm{gw}$ as the two-point correlation function (2PCF) of the source anisotropies alone, $C_\mathrm{gw}\equiv\ev{\delta^{(\mathrm{s})}_\mathrm{gw}\delta^{(\mathrm{s})}_\mathrm{gw}}$.
If we now include the kinematic dipole and calculate the 2PCF of the total density contrast $\delta_\mathrm{gw}$, then we find
    \begin{align}
    \begin{split}
        \label{eq:total-correlation}
        &\ev{\delta_\mathrm{gw}\delta_\mathrm{gw}}=\ev{\qty(\delta^{(\mathrm{s})}_\mathrm{gw}+\mathcal{D}\vu*e_\mathrm{o}\vdot\vu*v_\mathrm{o})\qty(\delta^{(\mathrm{s})}_\mathrm{gw}+\mathcal{D}\vu*e\mathrlap{'}_\mathrm{o}\vdot\vu*v_\mathrm{o})}\\
        &=\ev{\delta^{(\mathrm{s})}_\mathrm{gw}\delta^{(\mathrm{s})}_\mathrm{gw}}+2\mathcal{D}\ev{\delta^{(\mathrm{s})}\vu*e_\mathrm{o}\vdot\vu*v_\mathrm{o}}+\mathcal{D}^2\ev{\qty(\vu*e_\mathrm{o}\vdot\vu*v_\mathrm{o})\qty(\vu*e\mathrlap{'}_\mathrm{o}\vdot\vu*v_\mathrm{o})}\\
        &\approx C_\mathrm{gw}+\mathcal{D}^2\ev{\qty(\vu*e_\mathrm{o}\vdot\vu*v_\mathrm{o})\qty(\vu*e\mathrlap{'}_\mathrm{o}\vdot\vu*v_\mathrm{o})}.
    \end{split}
    \end{align}
We have taken the cross-correlation term as being approximately zero, as we expect there to be negligible correlation between the kinematic and cosmological terms (equivalently, $\delta^{(\mathrm{s})}_\mathrm{gw}$ is expected to important only at smaller angular scales).
The latter term in Eq.~\eqref{eq:total-correlation} can be evaluated by choosing spherical polar co\"ordinates such that
    \begin{align*}
        \vu*v_\mathrm{o}&=\qty(\sin\theta_v,0,\cos\theta_v),\qquad\vu*e_\mathrm{o}=\qty(0,0,1),\\
        \vu*e\mathrlap{'}_\mathrm{o}&=\qty(\sin\theta_\mathrm{o}\cos\phi_\mathrm{o},\sin\theta_\mathrm{o}\sin\phi_\mathrm{o},\cos\theta_\mathrm{o}).
    \end{align*}
Using the usual two-sphere metric to average over $\theta_v$ and $\phi_\mathrm{o}$ while keeping $\theta_\mathrm{o}$ fixed gives
    \begin{align*}
        &\ev{\qty(\vu*e_\mathrm{o}\vdot\vu*v_\mathrm{o})\qty(\vu*e\mathrlap{'}_\mathrm{o}\vdot\vu*v_\mathrm{o})}\\
        &\qquad\qquad=\ev{\cos\theta_v\qty(\cos\theta_v\cos\theta_\mathrm{o}+\sin\theta_v\sin\theta_\mathrm{o}\cos\phi_\mathrm{o})}\\
        &\qquad\qquad=\frac{1}{3}\cos\theta_\mathrm{o},
    \end{align*}
so that Eq.~\eqref{eq:total-correlation} therefore becomes
    \begin{equation}
        \ev{\delta_\mathrm{gw}\delta_\mathrm{gw}}\approx C_\mathrm{gw}+\frac{1}{3}\mathcal{D}^2\cos\theta_\mathrm{o}.
    \end{equation}
If we now let $\tilde{C}_\ell$ denote the modified $C_\ell$ that include the kinematic dipole, then we find
    \begin{equation}
        \tilde{C}_\ell=C_\ell+\frac{4\uppi}{9}\mathcal{D}^2\delta_{1\ell},
    \end{equation}
    where we have used the orthogonality property Eq.~\eqref{eq:legendre-orthogonal} and the fact that $P_1\qty(x)=x$.
Unsurprisingly, only the dipole component $C_1$ is affected.

\bibliography{gw-anisotropies}{}

\end{document}